\documentclass[11pt]{article}

\usepackage{geometry,setspace}
\small\normalsize

\usepackage{hyperref}
\hypersetup{
    colorlinks=true,
    citecolor=teal,
    linkcolor=teal
}
\usepackage{caption}
\usepackage{enumerate}
\usepackage{graphicx}
\usepackage{rotating,multirow}
\usepackage[table]{xcolor}
 \usepackage{amsmath,amsfonts,amssymb,mathrsfs,amsthm}
 \usepackage{mathtools}
\usepackage{bbm}
\usepackage{bm}
\usepackage[authoryear, round]{natbib}
\usepackage{booktabs}
\usepackage{pdflscape}
\usepackage{todonotes}
\usepackage{authblk}
\usepackage{xr}
\usepackage{tikz-cd}
\usepackage{framed}

\usepackage{xfrac}

\usepackage{caption}
\usepackage[T1]{fontenc}
\input{glyphtounicode}
\newtheorem{proposition}{Proposition}
\newtheorem{theorem}{Theorem}

\theoremstyle{definition}
\newtheorem{definition}{Definition}
\newtheorem{remark}{Remark}
\newtheorem{example}{Example}

\newcommand{\indicator}[1]{\ensuremath{I_{\{#1\}}}}

\usepackage{accents}
\newcommand{\vtrans}{\ensuremath{\mathcal{V}}}

\newcommand{\downprob}{\ensuremath{\delta}}

\newcommand{\Downprob}{\ensuremath{\Delta}}

\renewcommand{\P}{\mathbb{P}}
\newcommand{\E}{\mathbb{E}}
\newcommand{\R}{\mathbb{R}}

\newcommand{\Z}{\mathbb{Z}}

\newcommand{\eqdis}{\stackrel{\text{\tiny d}}{=}}

\newcommand{\var}{\operatorname{var}}
\DeclareMathOperator{\VaR}{VaR}

\renewcommand{\leq}{\leqslant}
\renewcommand{\geq}{\geqslant}

\renewcommand{\ge}{\geqslant}
\newcommand{\rd}{\mathrm{d}}

\newcommand{\svtrans}{\bm{\vtrans}}

\newcommand{\ourThanks}{\thanks{Acknowledgement. Initial ideas were
    developed while AJM was a guest of the Institute for
    Mathematical Research (FIM) at ETH Zurich. MB would like to acknowledge financial support from the Swiss National Science Foundation Project 200021\_191984.}}

\begin{document}
\title{Time series copula models using d-vines and
  v-transforms}
 \author{Martin Bladt}
 \affil{University of Lausanne}
 \author{Alexander J.\ McNeil\ourThanks}
 \affil{The York Management School, University of York}
\date{13 July 2021}
 \maketitle
 \begin{abstract}
An approach to modelling volatile financial return series using
stationary d-vine copula processes
combined with Lebesgue-measure-preserving transformations known as v-transforms is proposed. By
developing a method of stochastically inverting v-transforms, models
are constructed that can
describe both stochastic volatility in the magnitude of price
movements and serial correlation in their
directions. In combination with parametric marginal
distributions it is shown that these models can rival and sometimes outperform
well-known models in the extended GARCH family.
\end{abstract}
\section{Introduction}

The concept of a v-transform~\citep{bib:mcneil-20} facilitates the
application of copula models to
time series where the dominant feature is stochastic volatility, such as
financial asset return series. In the copula modelling approach to a single time
series $\{x_1, \ldots, x_n\}$ the idea is to find an appropriate strictly
stationary stochastic process $(X_t)$
consisting of a continuous marginal distribution $F_X$ and a copula process
$(U_t)$, given by $U_t = F_X(X_t)$ for all $t$, which models the serial dependencies in the
data; the latter is a process of standard uniform variables with
higher-dimensional marginal distributions that may be described by a
family of copulas $C_{\bm{U}}(u_1,\ldots,u_d)$ for $d\geq 2$.

For volatile return data, it is well known that serial dependence
becomes more
apparent under transformations like the absolute-value transformation
$T(x) = |x|$ or the
squared-value transformation $T(x) = x^2$, which remove directional information and
summarise the magnitude of price movements in what we term a volatility proxy time
series.

In \citet{bib:mcneil-20} general asymmetric volatility proxy transformations
$T(x)$ with change points $\mu_T$ are considered; these are continuous
functions which are increasing in $(x-\mu_T)$
for $x > \mu_T$ and increasing in $(\mu_t -x)$ for $x \leq
\mu_T$. Under such transformations,
the relationship between the terms of the copula process
$(U_t)$ of $(X_t)$ and the terms of the copula process $(V_t)$ of the volatility
proxy process $(T(X_t))$ 
can be described by a v-shaped function, known as a v-transform, which
is a mapping $\vtrans:[0,1] \to [0,1]$ that preserves the
uniformity of uniform random variables. The relationships between the
transformations are shown in diagram~\eqref{eq:A}.
\begin{equation}\label{eq:A}
\begin{tikzcd}[column sep=large]
  X_t \ar{r}{F_X} \ar{d}{T} & U_t \ar{d}{\vtrans}  \\
  T(X_t)  \ar{r}{F_{T(X)}} & V_t
\end{tikzcd}
\end{equation}

The key idea is that, rather than modelling serial dependence of the
time series at the level of $(U_t)$, we can model it at the level of
$(V_t)$ and create a composite copula model consisting of a family of copulas
$C_{\bm{V}}(v_1,\ldots,v_d),\; d=2,\ldots,n$ and a v-transform
$\vtrans$. In \citet{bib:mcneil-20} 
$C_{\bm{V}}$ is modelled using the implied copula process of an ARMA
model while in this paper we apply d-vine copula
models~\citep{bib:aas-czado-frigessi-bakken-09,
  bib:smith-min-almeida-czado-10}.

The resulting copula models, when combined with suitable marginal
distributions, yield an extremely flexible class of non-linear time
series models for volatile data.
In common with the popular GARCH family~\citep[among others]{bib:engle-82,bib:bollerslev-86,bib:ding-engle-granger-93,bib:glosten-jagannathan-runkle-93} and models from the more
general GAS (generalized autoregressive score) family~\citep{bib:creal-koopman-lucas-13}, the resulting
models are observation-driven volatility models, but they employ a
rather different mechanism in which the key feature is the strict
separation of marginal and serial dependence modelling. In the
GARCH paradigm the mechanism employed to capture serial
dependence behaviour can have a
constraining effect on the resulting marginal behaviour,
which is only partly mitigated by varying the innovation
distribution. For example, the standard GARCH model yields
marginal distributions with tails following power
laws for a wide class of innovation distributions~\citep{bib:mikosch-starica-00} and we will show in our
examples that these are not
always appropriate for observed return series.

There is a large literature on copula models for time series and good
starting points are the comprehensive review papers 
by~\citet{bib:patton-12} and~\citet{bib:fan-patton-14}. While the main focus
of this literature has been on cross-sectional dependence between multiple time
series, there is also a growing literature on modelling serial
dependence within single series and lagged dependence across series.
Markov copula models were first investigated
by~\cite{bib:darsow-nguyen-olsen-92} and
are further studied
in~\citet{bib:chen-fan-06b}. In the latter paper and
in~\cite{bib:chen-wu-yi-09} the theory of
semi-parametric estimation for these models is developed,
while~\cite{bib:beare-10} studies mixing properties of the resulting
processes; an application to data is given
in~\cite{bib:domma-giordano-perri-09}. Although theoretically
interesting, first-order Markov models are not realistic candidates
for modelling the persistent dependence and stochastic volatility 
that is found in typical
financial return series.

A distinct approach to copula modelling in statistics uses pair-copula
constructions. A key reference for applications in risk modelling is~\cite{bib:aas-czado-frigessi-bakken-09}, which builds on underpinning
work on joint density decompositions by~\cite{bib:joe-96,bib:joe-97}
and on graphical dependence models
by \cite{bib:bedford-cooke-01,
  bib:bedford-cooke-01b,bib:bedford-cooke-02} and
\cite{bib:kurowicka-cooke-06}. The
application of this methodology to modelling longitudinal dependence
in time series with d-vines was developed
in~\citet{bib:smith-min-almeida-czado-10} and the extension of this approach to
bivariate processes with both serial and cross-sectional dependence
using alternative vine structures is
treated
in~\cite{bib:beare-seo-15},~\cite{bib:brechmann-christian-czado-15},~\cite{bib:smith-15}
and~\cite{bib:nagler-krueger-min-20}.

The first-order Markov models
in~\citet{bib:chen-fan-06b} are special cases of the d-vine copula
approach of~\citet{bib:smith-min-almeida-czado-10}  but
the general d-vine pair-copula model allows higher-order Markov
dependence of the kind analysed in~\cite{bib:ibragimov-09}.
In applications of these models, the pair-copula building blocks used
by researchers have tended to be limited to a number of
well known bivariate copulas, such as the Gumbel, Clayton, Gaussian, t,
Frank and Joe copulas, as well as rotations of certain copulas through 90, 180
and 270 degrees. None of these basic copulas are particularly effective at
capturing the particular forms of serial dependence created by
stochastic volatility, in which large price movements are
followed be other large price movements, but of frequently changing sign.

\cite{bib:louaiza-maya-et-al-18} observe that these sign changes tend
to lead to lag-plots of log returns on the copula scale that are
cross-shaped. They address the shortcomings of standard pair
copulas in d-vine models by creating mixtures of pair copulas and rotated pair copulas
which can emulate these cross-shaped patterns.
In this paper we will show that the standard
copulas can be combined with v-transforms and d-vines to offer a parsimonious
method of obtaining a similar effect. Moreover the approach has more
econometric interpretability in that the driver of serial dependence is
identified with a volatility proxy series.



The contributions of the paper are threefold: we extend the theory of
v-transformed copula processes as presented in~\citet{bib:mcneil-20}
to allow models that can describe both the phenomenon of stochastic
volatility, as well as serial correlation in the direction of price
movements; we show how to apply the modelling framework to
copula processes based on d-vines and develop an approach to
estimation;
we demonstrate that the resulting models, when
combined with suitable marginal distributions, can rival and sometimes
outperform popular models in the GARCH class.

The paper is structured as follows. In
Section~\ref{sec:v-transforms-time} we extend the theory of copula
processes constructed using v-transforms; in particular, we explore a
generalization of the concept of
stochastic inversion of v-transforms. Section~\ref{section-Dvine}
shows how the theory applies to d-vine copula processes and
Section~\ref{sec:estimation} explains our approach to the estimation
of models and empirical examples
are presented in Section~\ref{sec:empirical-results}.
We apply the
fitted models to value-at-risk (VaR) estimation 
and analyse their out-of-sample forecasting performance in
Section~\ref{sec:using-model-forec};
Section~\ref{sec:conclusion} concludes.

\section{V-transforms and time series copula processes}\label{sec:v-transforms-time}
\subsection{V-transforms of uniform random variables}\label{sec:v-transforms-uniform}

In~\citet{bib:mcneil-20} three equivalent definitions of v-transforms
are provided. Suppose we consider absolutely continuous and strictly
increasing cdfs $F_X$ on $\R$ and volatility proxy transformations $T$
that are (i) continuous, (ii) strictly increasing for $x
\geq \mu_T$, (iii) strictly decreasing for $x  < \mu_T$ and (iv) differentiable
everywhere except at a change point parameter $\mu_T$ which may or
may not be zero; examples are $T(x)= |x|$ as well as alternatives that
are asymmetric around $\mu_T$. Then a 
v-transform
is as a function $\vtrans:[0,1] \to [0,1]$ constructed from $F_X$ and
$T$ by
\begin{equation}\label{eq:17}
  \vtrans(u) = F_{T(X)}\left( T\left( F_X^{-1}(u) \right) \right)
\end{equation}
where $F_{T(X)}$ denotes the cdf of $T(X)$ for any random variable $X$
with cdf $F_X$. Observe that the composite transformation in~\eqref{eq:17} represents
an excursion round three sides of the rectangle in diagram~\eqref{eq:A}, from
top right to bottom right.

$\vtrans$ is thus a mapping of the
probability-integral transform (PIT transform)
of $X$ to the PIT transform of $T(X)$ since $\vtrans( F_{X}(X)) =
F_{T(X)}(T(X))$. Clearly, by the properties of the PIT transform, such a transformation will preserve the
uniformity of uniform random variables: if $U \sim
U(0,1)$ and $V = \vtrans(U)$ then $V \sim U(0,1)$. More formally,
$\vtrans$ is a Lebesgue measure-preserving transformation on
the Borel subsets of $[0,1]$.

 A more visually interpretable definition is the following:
\begin{definition}\label{def:v-transform}
A v-transform is a mapping $\vtrans:[0,1] \to [0,1]$ with the
following properties:
  \begin{enumerate}
\item $\vtrans(0) = \vtrans(1) = 1$;
 \item There exists a point $\downprob$ known as the fulcrum
   such that $0 < \downprob < 1$ and $\vtrans(\downprob) = 0$;
 \item $\vtrans$ is continuous;
\item $\vtrans$ is strictly decreasing on
     $[0,\downprob]$ and strictly increasing on
       $[\downprob, 1]$;
       \item Every point $u \in [0,1]\setminus \{\downprob\}$ has a
         dual point $u^*$ on the opposite
         side of the fulcrum satisfying $\vtrans(u) = \vtrans(u^*)$
         and 
         $|u^* - u | = \vtrans(u)$ (square property).
\end{enumerate}
\end{definition}

Finally it is useful to have a definition which shows how v-transforms
can be easily constructed and this is afforded by the following characterization.

\begin{theorem}\label{theorem:v-characterization}
A mapping $\vtrans: [0,1] \to [0,1]$ is a v-transform  if and only if it takes the form
  \begin{equation}
    \label{eq:1_v}
    \vtrans(u) =
    \begin{cases}
(1-u) - (1-\downprob) \Psi \left( \frac{u}{\downprob} \right) & u \leq
\downprob, \\
u - \downprob \Psi^{-1}\left( \frac{1-u}{1-\downprob} \right) & u > \downprob,
\end{cases}
\end{equation}
where $\Psi$ is a continuous and strictly increasing distribution
function on $[0,1]$.
\end{theorem}
Parametric families of v-transforms may be obtained by assuming, for
example, that $\Psi$ is the cdf of a beta distribution or $\Psi(x) = \exp(-\kappa(-
(\ln x)^\xi))$ for $\kappa >0$ and $\xi >0$, which is
the main family considered in~\citet{bib:mcneil-20}. Both families include
the important special case of the linear v-transform 
\begin{equation}
  \label{eq:2_v}
 \vtrans_\downprob(u) =
\begin{cases}
 (\downprob-u)/\downprob& \quad u \leq \downprob,\\
 (u-\downprob)/(1-\downprob)& \quad u > \downprob,
\end{cases}
\end{equation}
which corresponds to a uniform cdf for $\Psi$ and which subsumes the symmetric
case $\vtrans_{0.5}(u) = |2u -1|$.


If we write a volatility proxy transformation in the form
\begin{align*}
T(x)=\left\{\begin{array}{ll}T_{1}\left(\mu_{T}-x\right) & x \leqslant \mu_{T} \\ T_{2}\left(x-\mu_{T}\right) & x>\mu_{T}\end{array}\right.
\end{align*}
for strictly increasing and continuous $T_1$ and $T_2$ satisfying $T_1(0) = T_2(0)$,
then it can be shown that the v-transform $\mathcal{V}$ in~\eqref{eq:17} is
determined by $F_X$, the value $\mu_T$ and the \textit{profile function} $g_{T}(x)=T_{2}^{-1} \circ
T_{1}(x)$. Different volatility proxy transformations
  may share the same change point $\mu_T$ and profile function $g_T$.
  For a fixed
distribution $F_X$ the pair $(\mu_T,g_T)$ partitions the
set of volatility proxy transformations into equivalence classes,
each corresponding to a unique v-transform\footnote{For example, the
  volatility proxy transformations $T(x) = |x|$, $T(x) = x^2$ and
  $T(x) = \ln|x|$ are all in the same equivalence class described by
  $\mu_T=0$ and $g_T(x) = x$.}. From this point of view,
having selected $F_X$, the selection of a v-transform
amounts to an implicit choice of a class of volatility proxy
transformations.

\begin{figure}[htb]
  \centering
   \includegraphics[width=16cm,height=7cm]{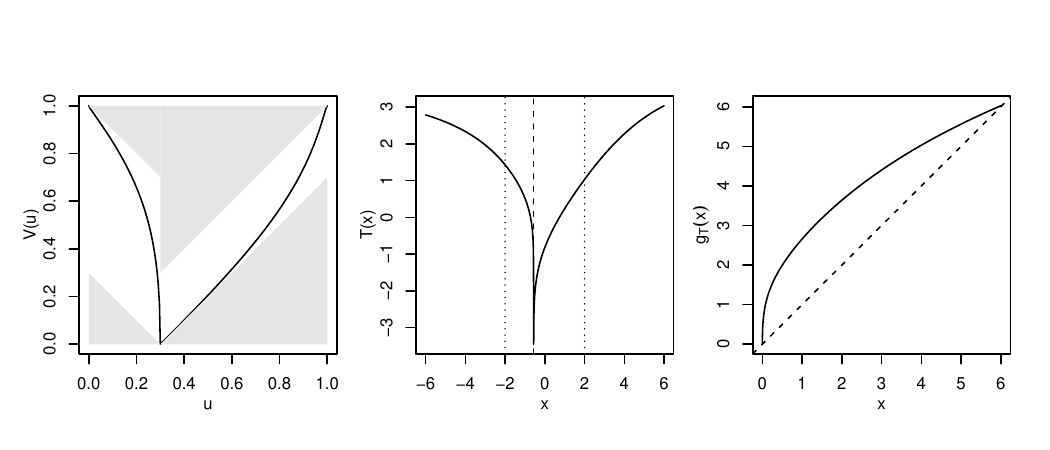}
   \caption{\label{fig:vt}Illustrations of the
     v-transform $\vtrans$ the
implied volatility proxy transformation $T$ and the profile function
$g_T$ in Example~\ref{example:vt}.}
 \end{figure}

\begin{example}\label{example:vt}
  Let $F_X$ be the distribution function of a standard Student t
  distribution with 5 degrees of freedom.
Consider the v-transform $\vtrans$ obtained by using the
generator $\Psi(x) = \exp(-\kappa(-
(\ln x)^\xi))$ in~(\ref{eq:1_v})  with parameter choices $\delta
=0.3$, $\kappa = 2.5$ and $\xi =0.5$.

The left panel of Figure~\ref{fig:vt} shows $\vtrans$ together with the admissible area (in
white) corresponding to the fulcrum value $\delta=0.3$; the
restriction arises from the aforementioned square property. The centre
and right panels show
the
implied volatility proxy transformation $T(x) =
\Phi^{-1}(\vtrans(F_X(x)))$, where $\Phi$ is the standard normal df,  and the profile
function $g_T$; note that $T$ is just one possible member of the
equivalence class defined by $F_X$ and $\vtrans$. The changepoint
$\mu_T = F_X(\delta)$ is negative and is marked by a dashed vertical line; two
further dotted
vertical lines show that an $X$ value at -2 is associated with
higher volatility than a value at 2.
The profile function is determined by~\citep[see][Proposition~3]{bib:mcneil-20}
\begin{align*}
g_{T}(x)=F_{X}^{-1}\left(F_{X}\left(\mu_{T}-x\right)+\mathcal{V}\left(F_{X}\left(\mu_{T}-x\right)\right)\right)-\mu_{T}, \quad x \geqslant 0.
\end{align*}
The dashed line shows the profile function for a
symmetric volatility proxy transformation. In this example, the shape
of the curve
$g_T(x)$ indicates that, for $x\geq 0$, a realization at $\mu_T-x$ has a
greater effect on volatility than a realization at $\mu_T+x$, but that
this effect wears off for larger $x$.
\end{example}

Note that, as the fulcrum parameter $\delta \to 0$ in~\eqref{eq:1_v} the
v-transform tends towards the function $\vtrans_0(u) = u$ on $(0,1]$
and as $\delta \to 1$ it tends towards the function $\vtrans_1(u) =
1-u$ on $[0,1)$. We refer to $\vtrans_0 : [0,1] \to [0,1],
\vtrans_0(u) = u$ and $\vtrans_1 : [0,1] \to [0,1],
\vtrans_1(u) = 1-u$ as degenerate v-transforms; the former corresponds
to any strictly increasing transformations of a continuously
distributed random variable $X$ and the latter to any
strictly decreasing function.

\subsection{Stochastic inversion of a v-transform}

Our aim is to develop stochastic processes for the process of uniform
random variables $(V_t)$ depicted in diagram~\eqref{eq:A}, and hence
to build processes $(X_t)$ to model financial returns. To this end we
need to be able to invert a v-transform $\vtrans$, but this is complicated by
the fact that $\vtrans$ is not one-to-one.
Stochastic inversion refers to the
process of randomly reversing a v-transform to arrive back at one of
the two dual points that yield the same value. To
develop stochastic processes for $(X_t)$ in which we have control over
the marginal distribution, we need to be able to do this in such  a way that uniformity is 
preserved under the stochastic inversion.

We introduce some further notation. Let $\vtrans^{-1}$ denote the
partial inverse given by
$\vtrans^{-1}: [0,1] \to [0,\delta], \;
\vtrans^{-1}(v) = \inf\{u :\vtrans(u) = v\}$ and let $\vtrans^\prime$
denote the gradient of $\vtrans$. The gradient of a
v-function is defined
for all points $u \in [0,1] \setminus \{\downprob\}$ and we adopt the
convention that $\vtrans^\prime(\downprob)$ is the left derivative as
$u \to \downprob$.

If two uniform random variables are linked by the v-transform $V =
\vtrans(U)$ then the joint distribution function of $(U,V)$ is a
special kind of copula. \cite{bib:mcneil-20} showed that, conditional on $V
= v$,
\begin{equation}\label{eq:35}
  U =
  \begin{cases}
    \vtrans^{-1}(v) & \text{with probability $\Delta(v),$} \\
    v +  \vtrans^{-1}(v) & \text{with probability $1-\Delta(v),$}
    \end{cases}
  \end{equation}
    where the function
\begin{equation}
  \label{eq:7}
  \Downprob(v) = - \frac{1}{\vtrans^\prime(\vtrans^{-1}(v))} \,.
\end{equation}
is referred to as the conditional down
probability of the v-transform and satisfies
$\E\left(\Downprob(V)\right) = \downprob$. This allows the concept of
\textit{the stochastic inversion function} of
a v-transform to be defined. This is simply a function that
facilitates the construction of a Bernoulli event by which a value of
$V$ is randomly assigned to one of the dual points $U$ and $U^*$ such
that $\vtrans(U) = \vtrans(U^*) = V$.
\begin{definition}[Stochastic inversion function of a
  v-transform]\label{def:stochinverse}
  Let
$\vtrans$ be a v-transform with conditional down probability
$\Delta(\cdot)$. The two-place function
$\bm{\vtrans}^{-1} : [0,1] \times [0,1] \to [0,1]$ defined by
\begin{equation}\label{eq:1}
  \bm{\vtrans}^{-1}(v,w) = 
  \begin{cases}
    \vtrans^{-1}(v) & \text{if $w \leq \Delta(v)$} \\
v + \vtrans^{-1}(v) & \text{if $w > \Delta(v)$.}
\end{cases}
\end{equation}
is the stochastic inversion function of $\vtrans$.
\end{definition}
It is obviously true that $\vtrans(\svtrans^{-1}(v,w)) = v$ for any $w$.
It is also simple to show that if $V$ and $W$ are independent $
U(0,1)$ random variables and $U = \svtrans^{-1}(V,W)$, then $U \sim
U(0,1)$. This is because
\begin{equation}\label{eq:6}
  \P\left(s\vtrans^{-1}(V,W) = \vtrans^{-1}(v) \mid V = v \right) =
  \P\left(W \leq \Delta(v) \mid V = v \right) = \P(W \leq \Delta(v) )= \Delta(v)
\end{equation}
so $U$ has the conditional distribution given in~(\ref{eq:35}) and must
be uniformly distributed.

When we apply a v-transform $\vtrans(u)$ followed by a stochastic inversion of the v transform, then we either arrive
back at the point $u$ or at its dual point $u^*$. In 
the next result we consider the sequence of uniformity-preserving
transformations
$U \to \vtrans(U)
\to \svtrans^{-1}(\vtrans(U),W)$ for $U$ and $W$ independent and
quantify the probability of arriving back at our starting point.
\begin{proposition}\label{prop:inverseplustransform}
Let $U \sim U(0,1)$ and $W \sim U(0,1)$ be independent random
variables and let $\vtrans$ be a v-transform with fulcrum $\downprob$
and conditional down probability $\Downprob(v)$. If $V =\vtrans(U)$ and
$\tilde{U} = \svtrans^{-1}(V,W)$ then
\begin{displaymath}
\P\left(\tilde{U} = U \right) 
= \downprob^2 + (1-\downprob)^2
+2\var(\Downprob(V)) \geq \downprob^2 + (1-\downprob)^2
\geq 0.5 \;.
\end{displaymath}
  \end{proposition}

We see that the probability $\P\left(\tilde{U} = U \right) $ that we recover the original value of $U$
is bounded below by
$\downprob^2 + (1-\downprob)^2 $. This value is attained for the linear
v-transform in~(\ref{eq:1}) since $\Delta(v) = \delta$ for all $v$ for
that family. The global minimum value is
$0.5$, which is attained only for the symmetric v-transform
$\vtrans_{0.5}$. Interestingly, when asymmetry is present, there is a
greater than 50\% chance of recovering the original value.

\subsection{V-transforms and inverse v-transforms of copulas}

V-transforms and their stochastic inversion functions can be applied componentwise
to random vectors. For vectors $\bm{u}$, $\bm{v}$ and $\bm{w}$ in $[0,1]^d$
we write
\begin{displaymath}
  \vtrans(\bm{u}) = (\vtrans(u_1),\ldots,\vtrans(u_d))^\prime \quad\text{and}\quad
  \bm{\vtrans}^{-1}(\bm{v},\bm{w}) =
(\bm{\vtrans}^{-1}(v_1,w_1), \ldots, \bm{\vtrans}^{-1}(v_d,w_d))^\prime
\end{displaymath}
for the componentwise operations.
 Let $\{(V_1,W_1),\ldots,(V_d,W_d)\}$ be a set of pairs of uniform random variables with the property
    that $V_i$ is independent of $W_i$ for all $i$. Note that these pairs need
    \textit{not} be independent of each other. Let $\bm{U} =
    \bm{\vtrans}^{-1} (\bm{V}, \bm{W})$ where $\bm{V} =
    (V_1,\ldots,V_d)^\prime$ and $\bm{W} =
    (W_1,\ldots,W_d)^\prime$. Then we know that:
    \begin{enumerate}
      \item
    $\bm{U}$ is a uniform random vector or, in other
    words, its joint distribution is a copula. This is guaranteed
    by the independence of $V_i$ and $W_i$ for all $i$ according
    to~\eqref{eq:6}.
    \item
      $\vtrans(\bm{U}) = \bm{V}$ regardless of the
    exact nature of the joint distribution of $(\bm{V},\bm{W})$ since
    $\vtrans(\bm{\vtrans}^{-1}(v,w)) = v$ for all $v,w$. It is possible to create different joint models
    $(\bm{V}_1,\bm{W}_1)$ and $(\bm{V}_2,\bm{W}_2)$ such that
    $\bm{V}_1 \eqdis \bm{V}_2$. In this case the implied copulas $\bm{U}_1 =
    \bm{\vtrans}^{-1} (\bm{V}_1, \bm{W}_1)$ and $\bm{U}_2 =
    \bm{\vtrans}^{-1} (\bm{V}_2, \bm{W}_2)$ are different but
    $\vtrans(\bm{U}_1) \eqdis \vtrans(\bm{U}_2)$.
  \end{enumerate}
  
    It is of interest to be able to determine the joint distribution
    of $\bm{U}$ under various assumptions about $(\bm{V},\bm{W})$.
    We give a result for the general case as well as the case
    where these vectors are independent of each other. This generalizes a result given
    in~\citet{bib:mcneil-20} for the case where $W_1,\ldots,W_d$ are
    also iid. 
To state this result compactly we introduce the notation
\begin{equation}\label{eq:3}
  \delta(u) = \Delta(
  \vtrans(u)), \quad
   I_{\delta,u} (x) =
  \begin{cases}
    [0, x] & u \leq \delta, \\
    [x, 1] & u > \delta,
  \end{cases}
  \quad
  p_{\delta,u} (x) =
  \begin{cases}
    x & u \leq \delta, \\
    1-x & u > \delta,
  \end{cases}
  \quad
  \delta \in (0,1),\quad u, x \in [0,1]
\end{equation}
and the vector form of the latter $p_{\delta,\bm{u}}(\bm{x}) =
    \left(p_{\delta,u_1} (x_1), \ldots, p_{\delta,u_d}
      (x_d)\right)^\prime$. 
Note that $\delta(u)$ is the
 probability that the v-transform of an observation at $u$ is
 assigned to the left side of the fulcrum under stochastic
 inversion. We now state the main result of this section.

\begin{theorem}\label{theorem:multivariate-vtransform}
  Let $\vtrans$ be a v-transform  and let $\{(V_1,W_1),\ldots,(V_d,W_d)\}$ be a set of pairs of uniform random variables with the property
    that $V_i$ is independent of $W_i$ for all $i$. Assume the copula
    of $(\bm{V},\bm{W})$ has a joint density.
Then the copula density
$c_{\bm{U}}(u_1,\ldots,u_d)$ of $\bm{U} = \bm{\vtrans}^{-1}(\bm{V},\bm{W})$ is 
\begin{equation}\label{eq:9}
  \frac{
  \int_{I_{\delta, u_1}(\delta(u_1))}\cdots
\int_{I_{\delta,
    u_d}(\delta(u_d))}c_{\bm{V},\bm{W}}\left(\vtrans(u_1),\dots,\vtrans(u_d),\,
  z_1,\dots,z_d\right) \rd z_1\cdots \rd z_d}{\prod_{i=1}^d  p_{\delta,u_i}\big(\delta(u_i)\big)}
\end{equation}
where $c_{\bm{V},\bm{W}}$ denotes the joint copula
density of $(\bm{V},\bm{W})$. When $\bm{V}$ and
$\bm{W}$ are independent this reduces to the simpler form

\begin{equation}\label{eq:15}
  c_{\bm{U}}(u_1,\ldots,u_d) =
  c_{\bm{V}}\big(\vtrans(u_1),\ldots,\vtrans(u_d)\big)
   \frac{C_{p_{\delta,\bm{u}}(\bm{W})}\Big(p_{\delta,u_1}\big(\delta(u_1)\big),\ldots,
    p_{\delta,u_d}\big(\delta(u_d)\big)\Big)}{\prod_{i=1}^d
    p_{\delta,u_i}\big(\delta(u_i)\big)}
\end{equation}
where $c_{\bm{V}}$ denotes the copula density of $\bm{V}$ and
$C_{p_{\delta,\bm{u}}(\bm{W})}$ denotes the copula of
$p_{\delta,\bm{u}}(\bm{W})$. When, in addition, $W_1,\ldots, W_d$
are independent,~\eqref{eq:15} reduces further to
$c_{\bm{V}}(\vtrans(u_1),\ldots,\vtrans(u_d))$.
\end{theorem}

\begin{figure}[h]
\centering
\includegraphics[width=5cm,trim=20cm 1cm 20cm 1cm,clip]{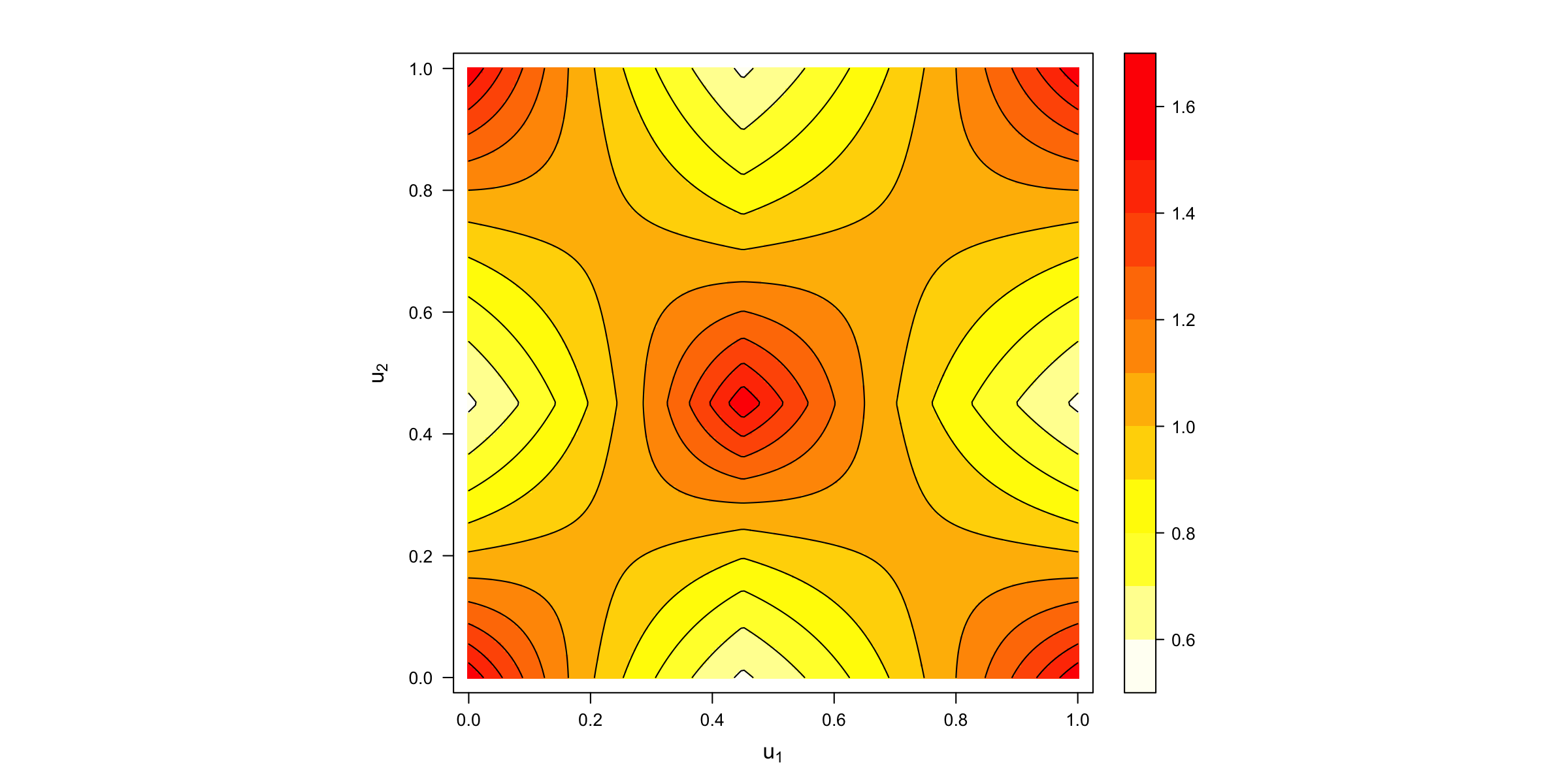}
\includegraphics[width=5cm,trim=20cm 1cm 20cm 1cm,clip]{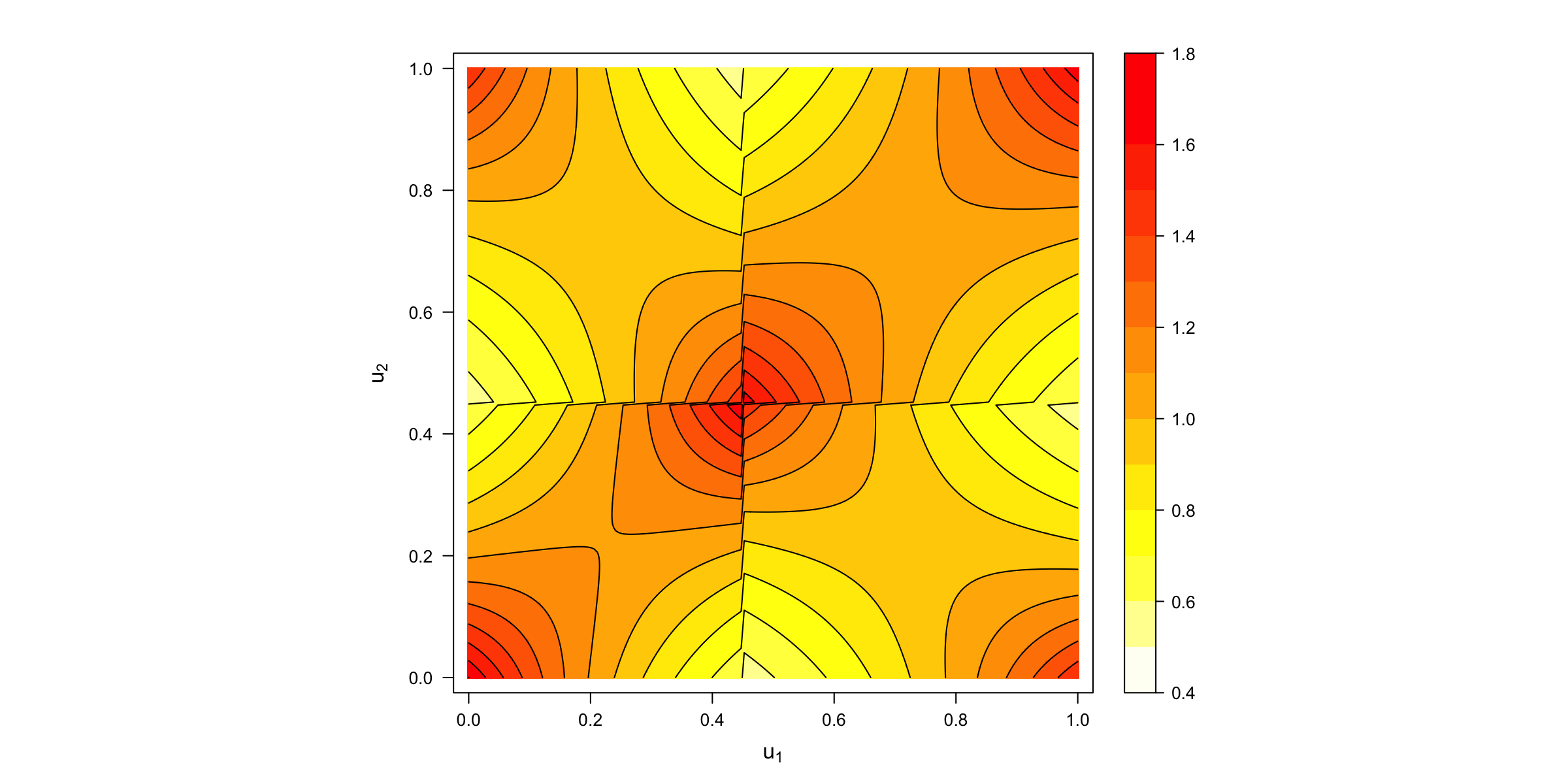}
\caption{Contour plot of
  $c_{\bm{V}}(\mathcal{V}(u_1),\mathcal{V}(u_2))$ (left) and
  $c_{\bm{U}}(u_1,u_2)$ (right) for model in which $\bm{V}$ follows
  Frank(1) copula and $\bm{W}$  follows Frank($0.5$) copula; v-transform is
  linear with fulcrum $\delta=0.45$.}
\label{V_GV_copulas}
\end{figure}

In the applied sections of this paper the focus will be
 on models of type~(\ref{eq:15}) and it is instructive to consider the
 structure of the copula in more detail. For $d=2$, Figure
 \ref{V_GV_copulas} illustrates
 $c_{\bm{V}}(\mathcal{V}(u_1),\mathcal{V}(u_2))$ and
 $c_{\bm{U}}(u_1,u_2)$ for particular choices of parametric copulas
 for $\bm{V}$ and $\bm{W}$ and for the linear v-transform. We observe
 the characteristic cross-shape often observed in lag-plots for
 processes with stochastic volatility.

\begin{figure}[h]
\centering
\includegraphics[width=5cm,trim=20cm 1cm 20cm 1cm,clip]{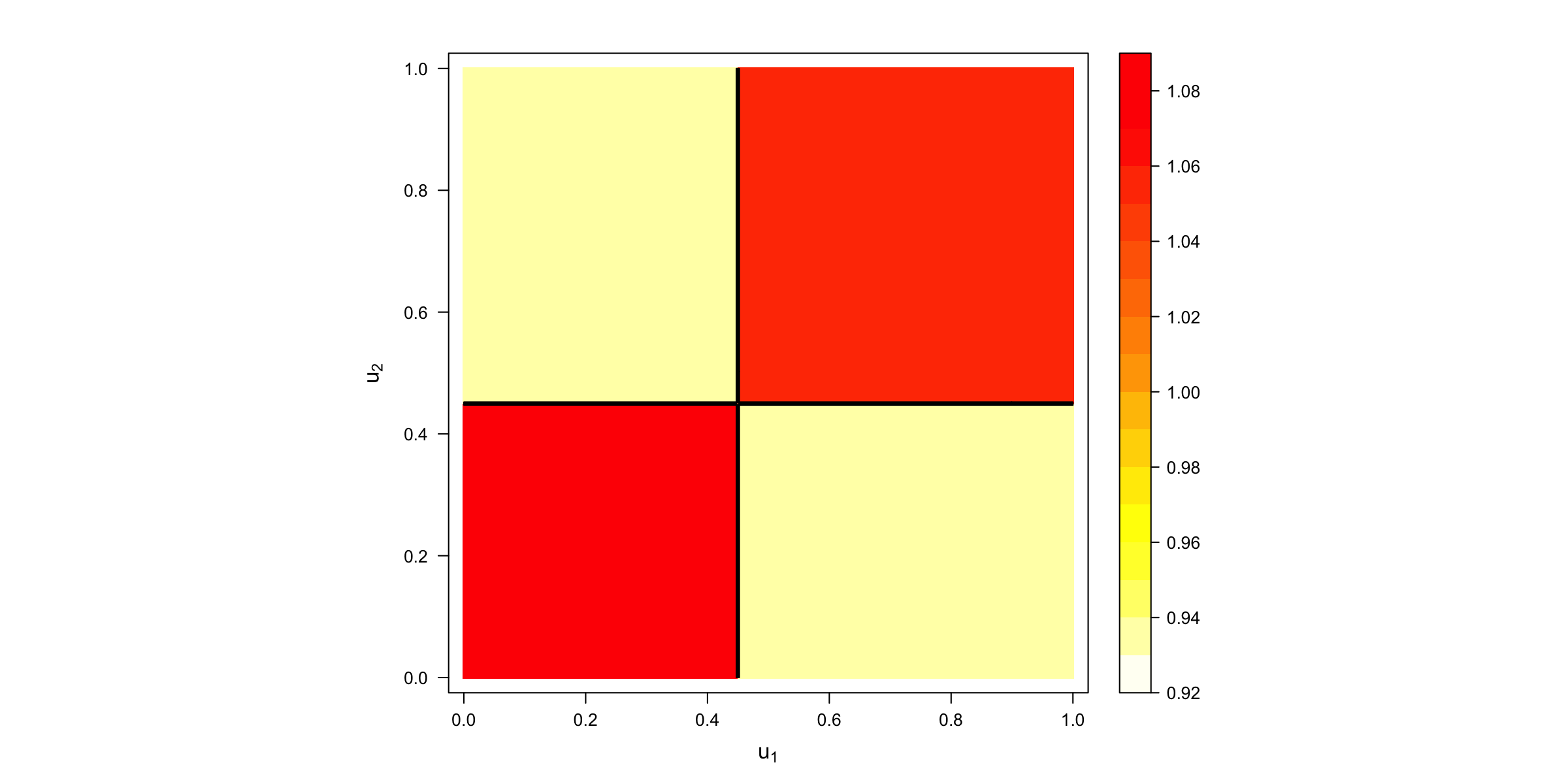}
\includegraphics[width=5cm,trim=20cm 1cm 20cm 1cm,clip]{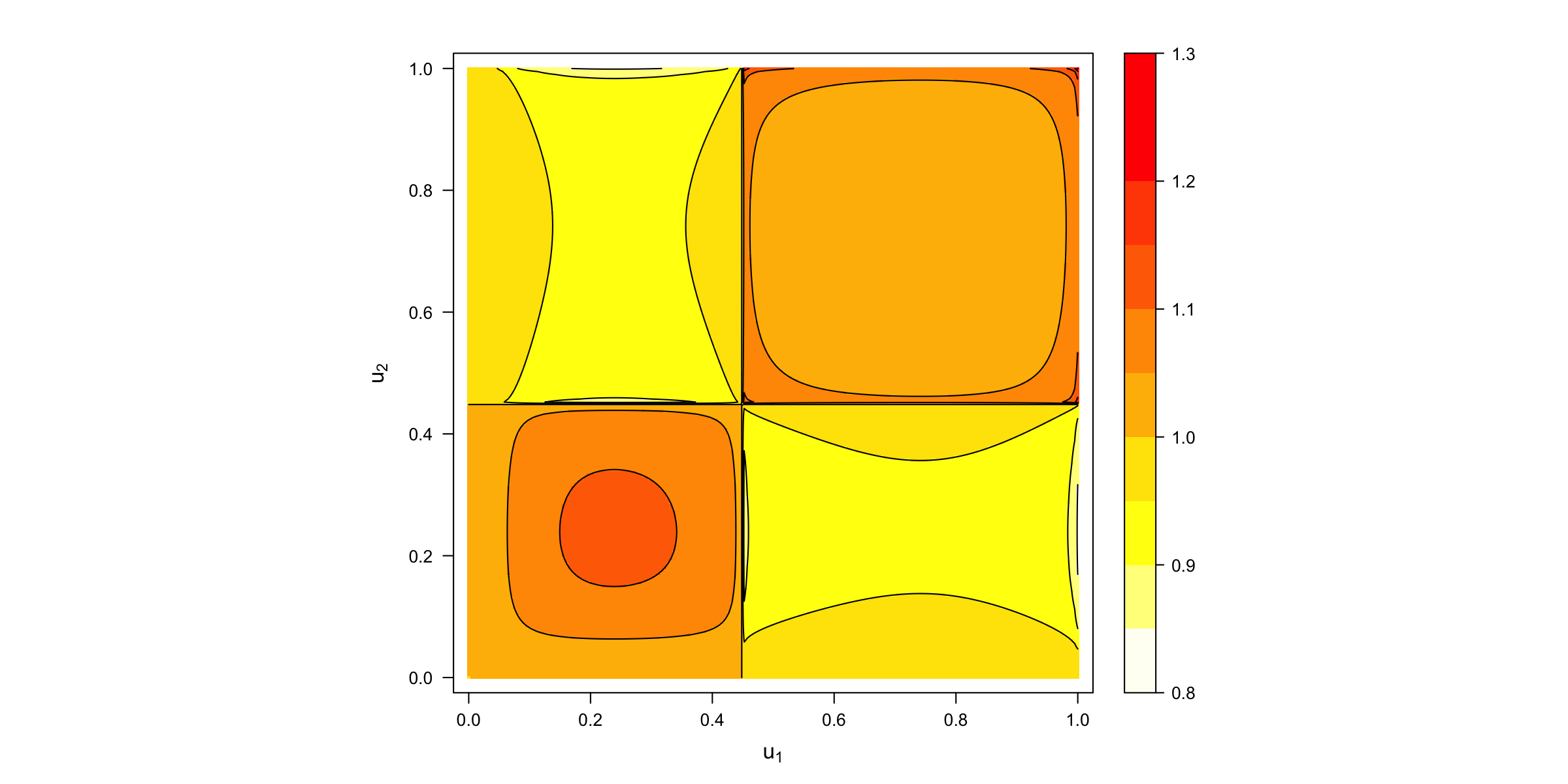}
\includegraphics[width=5cm,trim=20cm 1cm 20cm 1cm,clip]{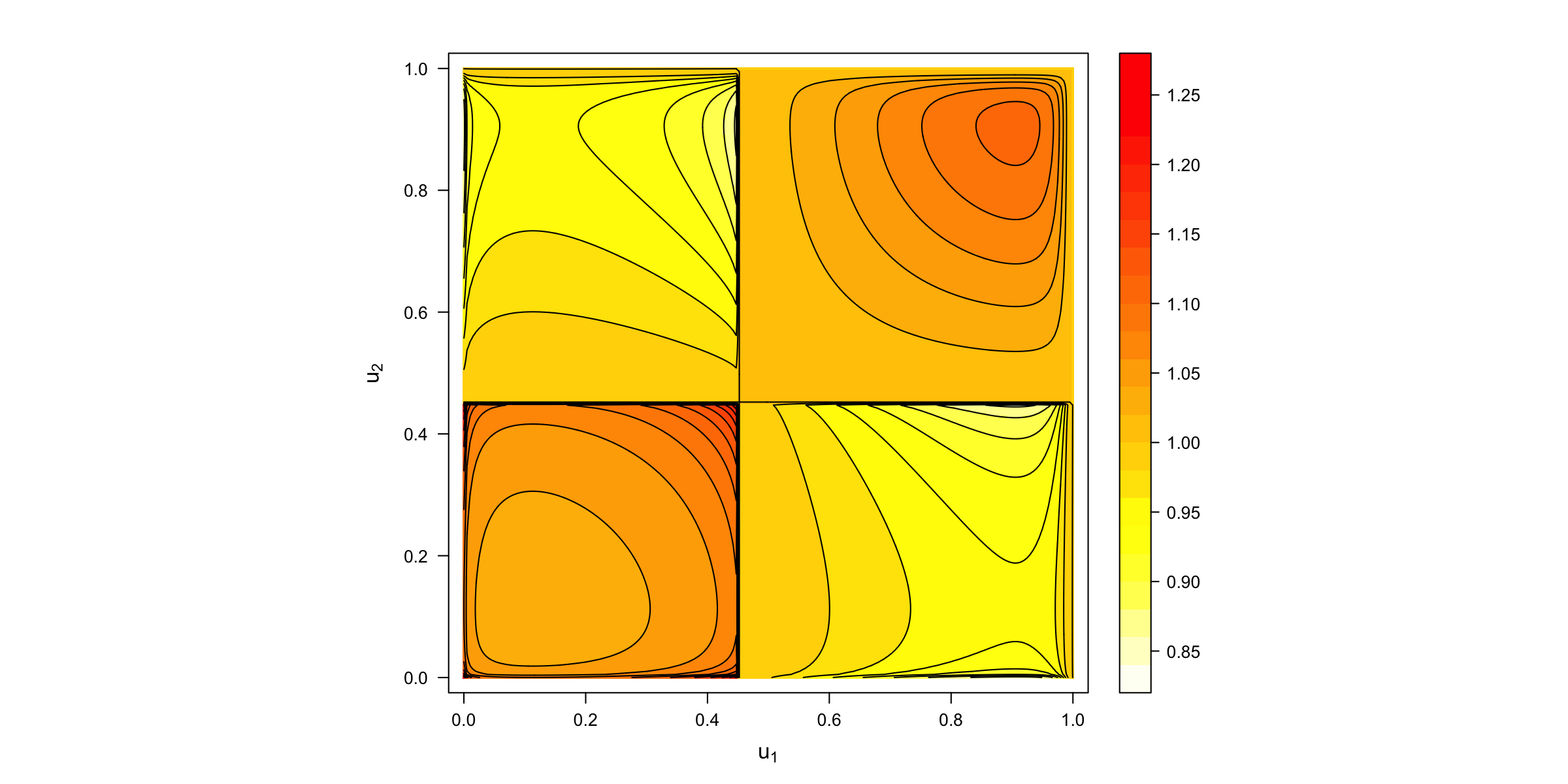}
\caption{Copula density $c_{\bm{W^\ast}}(u_1,u_2)$ when $\bm{W}$
  follows Frank($0.5$) copula for three different v-transforms, all with fulcrum $\delta=0.45$: 
linear (left); $\kappa=\xi=1.5$ (middle); $\kappa=1.5$, $\xi=0.5$ (right).}
\label{Wast_copulas}
\end{figure}

  The density in~(\ref{eq:15})
 is itself
 the product of two copula densities,
 $c_{\bm{V}}(\vtrans(u_1),\ldots,\vtrans(u_d))$ and the 
 density
 \begin{equation}\label{eq:2}
   c_{\bm{W}^\ast}(u_1,\ldots,u_d) :=  \frac{C_{p_{\delta,\bm{u}}(\bm{W})}\Big(p_{\delta,u_1}\big(\delta(u_1)\big),\ldots,
    p_{\delta,u_d}\big(\delta(u_d)\big)\Big)}{\prod_{i=1}^d
    p_{\delta,u_i}\big(\delta(u_i)\big)}\;.
 \end{equation}
 To see that this is a density observe that
 $c_{\bm{U}}(u_1,\ldots,u_d) = c_{\bm{W}^\ast}(u_1,\ldots,u_d)$
 when $\bm{V}$ is a vector of independent uniform
 variables so that $c_{\bm{V}}(v_1,\ldots,v_d) =1$.  In Figure
 \ref{Wast_copulas} we illustrate for three different choices of
 v-transform the copula density
 $c_{\bm{W^\ast}}(u_1,u_2)$, which controls the dependencies in
 directions of movements. 
 
\subsection{V-transforms of time series copula processes}

The theory presented in the previous section can obviously be applied to the
construction of time series copula processes $(U_t)_{t\in \Z}$ that are suitable
for modelling financial return data. By constructing a strictly stationary bivariate process $(V_t,W_t)_{t \in\Z}$
we obtain a strictly stationary process for $(U_t)$ through
the stochastic inversion construction of
Theorem~\ref{theorem:multivariate-vtransform}.  In these
models $(V_t)$ can be thought of as accounting mainly for
stochastic volatility (serial dependence in the magnitude of
movements) while $(W_t)$, if it is not an iid process, can account for extra serial dependence in the direction of price movements.

We restrict attention to the special case where $(V_t)$
and $(W_t)$ are independent processes as this results in the tractable joint density for
$(U_t)$ in~(\ref{eq:15}) and permits likelihood-based inference. Within this framework we consider models where $(W_t)$ is
either strict white noise (an iid process) 
or a first-order Markov process. In comparison with
stochastic volatility, dependence in the signs of asset returns is
a relatively weak and transient phenomenon and first-order Markov
models for $(W_t)$ appear to be sufficient in the majority of datasets
we have considered.

The process $(V_t)$ will be modelled by using a stationary d-vine
copula process of Markov order $k$ yielding the class of vt-d-vine models, which
complement the vt-ARMA class of models in~\citet{bib:mcneil-20}.
First-order Markov dependence in $(W_t)$ will be modelled using a d-vine
process of order $k = 1$, i.e. the kind of model considered
in~\cite{bib:chen-fan-06b}
and~\cite{bib:domma-giordano-perri-09}.

\section{D-vine and vt-d-vine copula processes}\label{section-Dvine}

The aim of this section is to develop copula
processes for $(V_t)$ and $(U_t)$ that fulfill the requirement of strict stationarity.
\subsection{D-vine copula processes}
Using the theory described in \cite{bib:smith-min-almeida-czado-10} the
multivariate copula
density $c_{\bm{V}}$ of a random vector $\bm{V} = (V_1,\ldots,V_d)^\prime$ can be decomposed as a d-vine taking the form
\begin{align}\label{D_vine_density}
  c_{\bm{V}}(v_1,\dots,v_d)&=
              \prod_{i=1}^{d-1}\prod_{t=1}^{d-i}
                  c_{t,t+i|S_{t,t+i}}(v_{t|S_{t,t+i}},v_{t+i|S_{t,t+i}}),
\end{align}
where $S_{t,t+i} =\{t+1,\ldots,t+i-1\}$ denotes the set of
indices of the variables lying between $V_t$ and $V_{t+i}$,
$c_{t,t+i|S_{t,t+i}}$ is a pair copula density (i.e.~a bivariate
copula density) describing the
dependence between variables $V_t$ and $V_{t+i}$ conditional on these
variables and
\begin{displaymath}\label{pseudo_cond}
v_{j|S_{t,t+i}}=\P(V_j \leq v_j|V_{t+1} = 
v_{t+1},\dots,V_{t+i-1} =
  v_{t+i-1}),\quad j \in \{t,t+i\}
\end{displaymath}
denotes the conditional cdf of variable $j$ conditional on the
intermediate variables; note that $S_{t,t+1} =
\emptyset$ and so the conditioning set is dropped in this case.

The decomposition~\eqref{D_vine_density} is
not the unique d-vine expression for $c_{\bm{V}}(v_1,\ldots,v_d)$ when $d>2$,
since the variables $v_1,\ldots,v_d$ could be arranged in other
orders. However, when the variables have a natural ordering, as they do
for a time series, then~\eqref{D_vine_density} is the canonical
expression for a d-vine. It should also be noted that there are other ways of decomposing a
joint density using pair copulas which are not d-vines but which belong
to the more general class of
regular vines investigated in~\cite{bib:bedford-cooke-01b}; we do not
consider these further as d-vines seem well adapted to the univariate
time series context.

A final important point is that the decomposition
\eqref{D_vine_density} of an arbitrary joint density may result in
pair copulas whose functional forms depend on the values of the
conditioning variables in the sets $S_{j,i}$. However, in applied statistics, \eqref{D_vine_density}
is used as a framework for constructing rather than deconstructing
models
and interest is usually confined to so-called \textit{simplified pair
copula constructions} in which the copula forms are invariant to the
values of the conditioning variables and are chosen from a number of
well-known parametric families. In this case we can simplify the
copula notation
to $c_{t,t+i} = c_{t,t+i\mid S_{t,t+i}}$. There has been quite a lot
of interest in the question of whether simplified constructions are
sufficiently flexible and robust to model all dependence structures;
see~\cite{bib:haff-aas-frigessi-10},~\cite{bib:stoeber-joe-czado-13},~\cite{bib:spanhel-kurz-19}
and~\cite{bib:mroz-fuchs-trutschnig-21}. While these authors draw
attention to limitations, the simplifying assumption still admits a rich
class of copulas which generalize the
dependence inherent in classical autoregressive time series models, as
we now explain.


It is possible to construct strictly stationary time series $(V_t)_{t \in \Z}$ in which
the $d$-dimensional joint densities of random vectors
$(V_{t+1},\ldots,V_{t+d})$ for $d\geq 2$ are given by the simplified
form of the decomposition 
\eqref{D_vine_density}. In this case the stationarity requirement
imposes the restriction that the pair copula densities $c_{t,t+i}$ may
only depend on $i$ and we can further simplify notation by writing $c_i = c_{t,t+i}$. Moreover,
by setting $c_i =1$ (corresponding to the independence copula) for
$i>k$ we obtain Markov processes of order $k$ and reduce the number of
copulas that need to be determined. Models of this kind are
investigated by \cite{bib:brechmann-christian-czado-15} (under the name
COPAR),~\cite{bib:smith-15} and \cite{bib:nagler-krueger-min-20}; in the latter paper it is shown that
d-vines are the only regular vines that can be used to construct
stationary univariate time series. 

\begin{definition}\label{definition:copar}
A time series $(V_t)_{t \in \Z}$ 
is a  d-vine($k$) copula process if, for any $d
\geq2$ and $t \in \Z$, the $d$-dimensional marginal
density of the random vector $(V_t,\ldots,V_{t+d-1})$ has the form
\begin{equation}
  \label{eq:13}
  c_{\bm{V}}(v_1,\dots,v_d) =
  \prod_{i=1}^{\min(k,d-1)} \prod_{t=1}^{d-i}c_i\big(v_{t \mid S_{t,t+i}}
, \;v_{t+i\mid S_{t,t+i}}\big)
\end{equation}
for a sequence of bivariate copula densities $c_1, \ldots,c_k$.
\end{definition}

A d-vine($k$) copula process is fully determined by the $k$ copula
densities corresponding to each of its conditional dependencies, or
\textit{generalized lags}. When $k=1$, the copula process reduces to that of \cite{bib:chen-fan-06b}. When
the copula densities are Gaussian, the dependence
structure is that of a Gaussian AR($k$) model, and the parameters of the pair
copulas in the d-vine model are the partial autocorrelations of the
underlying AR($k$) process. In general, for $k>1$, the main practical
difficulty lies in the calculation of the expressions $v_{t \mid
  S_{t,t+i}}$ and $v_{t+i\mid S_{t,t+i}}$ for $i >1$. This can be done
using the recursive identities
\begin{equation}\label{eq:18}
\begin{aligned}
  v_{t \mid S_{t,t+i}}
&= h_{i-1}^{(2)} \big(  v_{t \mid S_{t,t+i-1}} ,\;\;  v_{t+i - 1 \mid
                                                                       S_{t,t+i-1}} \big)
  \\
  v_{t+i\mid S_{t,t+i}}
&= h_{i-1}^{(1)} \big(   v_{t+1 \mid S_{t+1,t+i}},\;\;   v_{t+i \mid S_{t+1,t+i}}
                                                                                                 \big)
                                                                                               \end{aligned}
                                                                                               \end{equation}
where 
$h_i^{(j)}(v_1, v_2) = \frac{\partial}{\partial v_j} C_i(v_1, v_2), j
  \in \{1,2\}$,
denotes the partial derivative or $h$-function of the copula
$C_i$. Thus the problem is recursively reduced to the problem of
evaluating $h$-functions of bivariate copulas~\citep{bib:joe-96}. Note
that when $C_i$ is an exchangeable copula (satisfying $C_i(v_1,v_2) =
C_i(v_2, v_1)$ for all $v_1, v_2$) the calculation is further
simplified by the fact that $h_i^{(2)}(v_1, v_2) = h_i^{(1)}(v_2,
v_1)$ but when $C_i$ is non-exchangeable then both partial derivatives
must be calculated explicitly.

The d-vine process of Definition~\ref{definition:copar} is
strictly stationary by design but the questions of ergodicity and
mixing are
trickier. A number of authors
including~\cite{bib:chen-fan-06b},~\cite{bib:beare-10} and \cite{bib:longla-peligrad-12} have
studied the first-order process ($k=1$). \cite{bib:longla-peligrad-12}  show
that, if the density of the absolutely continuous part of a copula is strictly positive almost
everywhere, the resulting process is $\beta$-mixing (absolutely
regular) and therefore
ergodic. A variety of results have been obtained on the rate of mixing
and these typically depend on the tail dependence characteristics of
the copula. The Gauss and Frank copulas can be shown to be
$\phi$-mixing~\citep{bib:longla-peligrad-12} and thus geometrically $\beta$-mixing, while the Gumbel,
Clayton and t copulas satisfy the weaker property of geometric $\rho$-mixing~\citep{bib:beare-10}; see~\citet{bib:bradley-05} for more
details of mixing conditions.  \cite{bib:zhao-shi-zhang-18}~extend the Markov chain approach to models with order
$k>1$ and give general conditions for geometric ergodicity, but do not address
specific copula choices.

\subsection{Vt-d-vine copula processes}

\begin{definition}\label{def:vt-dvine}
Let $\vtrans$ be a v-transform, let $(V_t)_{t\in \Z}$ be a
d-vine($k$) copula process and let  $(W_t)_{t\in \Z}$ be any
strictly stationary copula process that is independent of $(V_t)$.
Let $(U_t)_{t \in \Z}$
be defined componentwise by setting $U_t = \bm{\vtrans}^{-1}(V_t,W_t)$.
\begin{enumerate}
  \item If $(W_t)$ is an iid process (strict white noise) we say that
  $(U_t)$ is a vt-d-vine($k$) copula process.
\item Otherwise $(U_t)$ is a  generalized
  vt-d-vine($k$) copula process, or a gvt-d-vine($k$) process.
\end{enumerate}
\end{definition}
It follows immediately
from Theorem \ref{theorem:multivariate-vtransform}, by inserting the
d-vine($k$) density~\eqref{eq:13} in the general
expression~(\ref{eq:15}) and using the
notation~(\ref{eq:2}), that the joint density of a gvt-d-vine($k$)
process is
\begin{equation}
  \label{eq:14}
  c_{\bm{U}}(u_1,\dots,u_d) =  c_{\bm{W}^\ast}(u_1,\ldots,u_d)
  \prod_{i=1}^{\min(k,d-1)} \prod_{t=1}^{d-i}  \left.  c_i\big(v_{t \mid S_{t,t+i}}
  , \;v_{t+i\mid S_{t,t+i}}\big) \right|_{v_1 =
  \vtrans(u_1),\ldots,v_d = \vtrans(u_d)}.
\end{equation}

 \begin{remark}
Under the degenerate v-transform $\vtrans_0$ with fulcrum set at zero we
have that
$\vtrans_0(u)=u$ for $u \in (0,1]$. In this case the joint density of
a vt-d-vine($k$) process
reduces to $c_{\bm{U}}(u_1,\ldots,u_d) = c_{\bm{V}}(u_1,\ldots,u_d)$ so
that a
d-vine($k$) copula process model can be considered as a boundary
case of a vt-d-vine($k$) model. 
\end{remark}

As noted earlier, we will use d-vine(1) models for the
process $(W_t)$. In this case the term 
$ c_{\bm{W}^\ast}(u_1,\ldots,u_d) $ in the joint density~(\ref{eq:14}) can be fully
expressed in terms of bivariate copula densities.
When $d=2$ let the joint distribution function of $
(W_1,W_2)$ be denoted $C_W= C_{\bm{W}}$ and let
$c_{W^\ast} = c_{\bm{W}^\ast}$. Furthermore let $C_{W^\prime}$, $C_{W^{\prime\prime}}$ $C_{W^{\prime\prime\prime}}$  denote
the joint cdfs of $(W_1,1-W_2)$, $(1-W_1,1-W_2)$ and $(1-W_1, W_2)$
respectively; these are the copulas
obtained by rotating the distribution described by $C_W$ through 90, 180
and 270 degrees clockwise. We then have the following expression for
$c_{\bm{W}^\ast}$.

\begin{proposition}\label{prop:Markov-conditioning}
When $(W_t)$ is first-order Markov
\begin{equation}\label{eq:10}
  c_{\bm{W}*}(u_1,\ldots,u_d)= \prod_{i=2}^d c_{W^\ast}(u_{i-1}, u_i),
  \; \;
c_{W^\ast}(u_1, u_2) =
 \begin{cases}
\frac{C_W\big(\delta(u_1),\delta(u_2)\big)}{\delta(u_1)\delta(u_2)}& u_{1} \leq
\delta, u_2 \leq \delta, \\
\frac{C_{W^\prime}\big(\delta(u_1),1-\delta(u_2)\big)}{\delta(u_1)(1-\delta(u_2))} & u_{1} \leq
\delta, u_2 > \delta, \\
\frac{C_{W^{\prime\prime}}\big(1-\delta(u_1),1-\delta(u_2)\big)}{(1-\delta(u_1))(1-\delta(u_2))} & u_{1} >
\delta, u_2 > \delta, \\
\frac{C_{W^{\prime\prime\prime}}\big(1-\delta(u_1),\delta(u_2)\big)}{(1-\delta(u_1))\delta(u_2)} & u_{1} >
\delta, u_2 \leq \delta.
\end{cases}                                                                                              
\end{equation}
\end{proposition}

The conditional density of the resulting gvt-d-vine($k$) copula
process can be calculated from~\eqref{eq:14} and~\eqref{eq:10}
and takes the form
\begin{align}\label{cond_dist_gvtcopar}
f_{U_t|U_{t-1},\ldots,U_{t-k}}(u_t|u_{t-1},\ldots,u_{t-k})&=
                                        \frac{c_{\bm{U}}(u_{t-k},\ldots,u_t)}{c_{\bm{U}}(u_{t-k},\ldots,u_{t-1})}\nonumber
  \\
  &=
 c_{W^\ast}(u_{t-1},u_t)
\prod_{i=1}^{k}  \left.  c_i(v_{t-i|S_{t-i,t}},v_{t|S_{t-i,t}}) \right|_{v_{t-k} =
  \vtrans(u_{t-k}),\ldots,v_t = \vtrans(u_t)}.
\end{align}
For $k=1$ this is simply
$f_{U_t|U_{t-1}}(u_t|u_{t-1})= c_{\bm{U}}(u_{t-1},u_t) =
 c_{W^\ast}(u_{t-1},u_t)c_1(\vtrans(u_{t-1}),\vtrans(u_t))$
which is the conditional density of a first-order Markov copula model with copula
$C_{\bm{U}}$. If the copula densities $c_1$ and $c_{W^\ast}$ are
strictly positive almost everywhere, the result
of~\cite{bib:longla-peligrad-12} implies the process is $\beta$-mixing
and ergodic. The requirements for mixing and ergodicity in the case
$k>1$ are an open question.

\begin{figure}[htb]
  \centering
   \includegraphics[width=14cm,height=6cm]{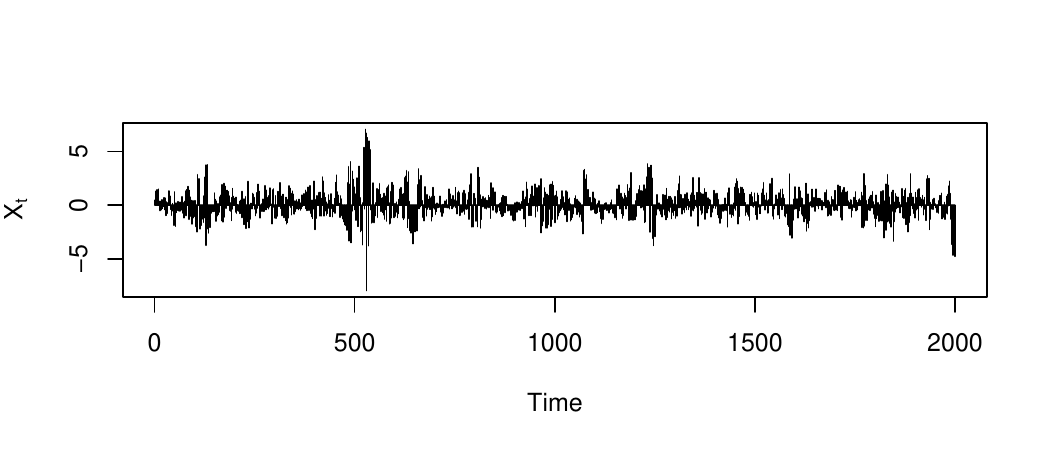}
   \caption{\label{sim_figA}Time series plot for a sample of size
     $n=2000$ from the process described in
     Example~\ref{example:simulated}.}
 \end{figure}

\begin{figure}[htb]
  \centering
   \includegraphics[width=14cm,height=10cm]{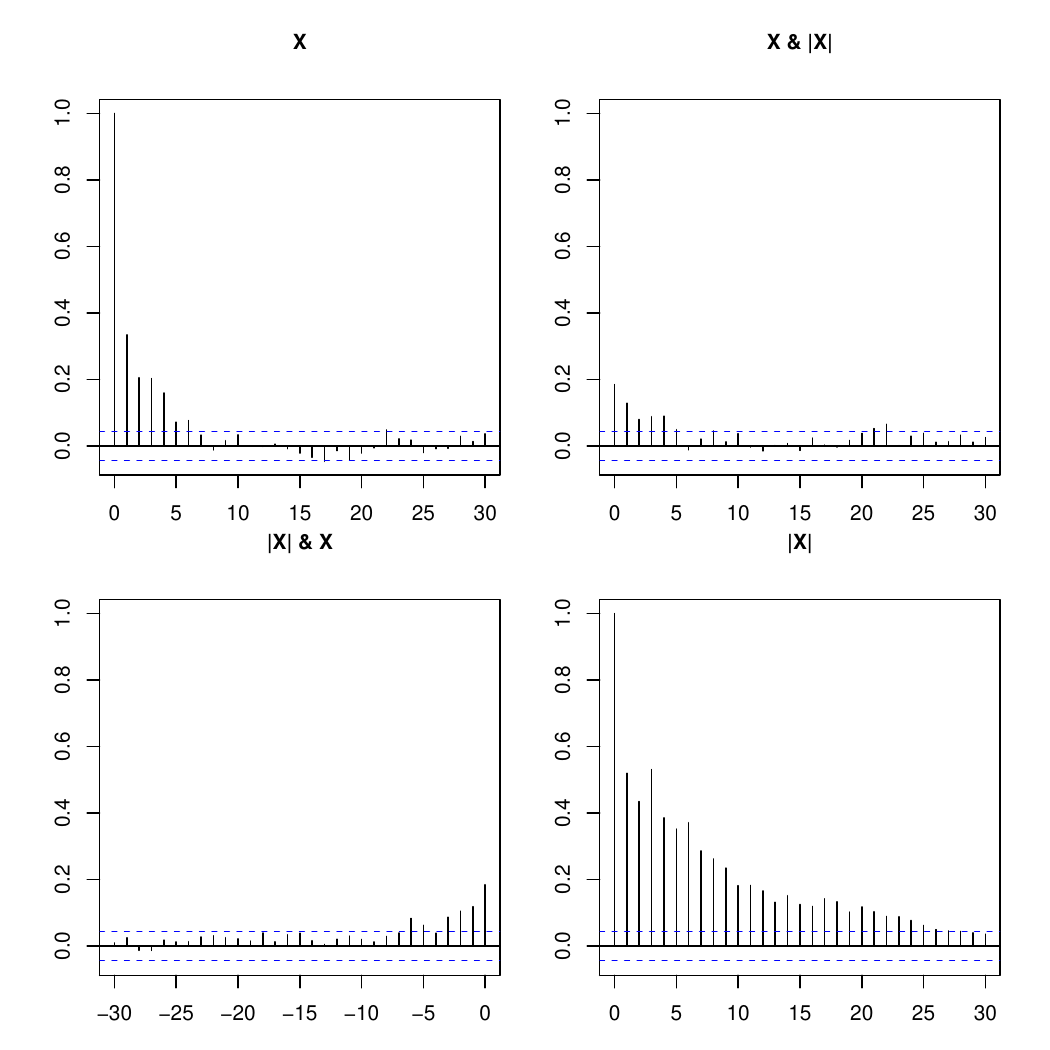}
   \caption{\label{sim_figB}Sample acf plots for a sample of size
     $n=2000$ from the process described in Example~\ref{example:simulated}.}
 \end{figure}

\begin{example}\label{example:simulated}
We construct a strictly stationary time series based on a gvt-d-vine(3) process with the follow
specification. The underlying d-vine(3) copula process has a
180-degree-rotated Clayton copula ($\theta=0.7$) at lag 1, a t copula
($\nu=5$ and $\rho=0.2$) at lag 2 and a Joe copula ($\theta=1.5$) at
lag 3. The v-transform  $\vtrans$ and marginal distribution $F_X$ are
as in Example~\ref{example:vt}; in particular, the marginal
distribution is Student t. The d-vine(1) copula process
for $(W_t)$ uses the Gaussian copula ($\rho = 0.7$). 

Figure~\ref{sim_figA} shows a realization of length $n=2000$
from the resulting process $(X_t)$ and Figure~\ref{sim_figB} shows that there is strong serial
dependence in $(|X_t|)$ as well as serial dependence in $(X_t)$ and between
$(X_t)$ and $(|X_t|)$.
\end{example}

    \section{Estimation}\label{sec:estimation}
We now turn to the statistical estimation of the processes defined in the previous section. Let $\bm{x} =
\{x_1,\ldots,x_n\}$ denote a realization from a strictly
stationary process with parametric marginal distribution
$F_X(x;\bm{\theta}_m)$ and joint copula density
$c_{\bm{U}}(u_1,\ldots,u_n;\bm{\theta}_c)$. The full log-likelihood is 
 \begin{align}\label{eq:24}
L(\bm{\theta}_c\, ,\bm{\theta}_m;\bm{x})=\log\left(c_{\bm{U}}(F_X(x_1;\bm{\theta}_m),\dots,F_X(x_n;\bm{\theta}_m);\bm{\theta}_c)\right)+\sum_{i=1}^n\log\left(f_X(x_i;\bm{\theta}_m)\right).
 \end{align}
 In copula-based inference it is common to build up the model
 componentwise by estimating marginal distribution and copula
  model in successive steps. We follow the inference-functions-for-margins (IFM)
  approach~\citep{bib:joe-97} in which we first maximize
  $L_1(\bm{\theta}_m;\bm{x})=\sum_{i=1}^n\log\left(f_X(x_i;\bm{\theta}_m)\right)$
  to obtain estimates of the marginal parameters $\bm{\widehat{\theta}}_m$ and then
  maximize
  $L_2(\bm{\theta}_c;\bm{u})=\log\left(c_{\bm{U}}(u_1\dots,u_n;\bm{\theta}_c)\right)$
  using pseudo-copula data $\{u_i = F_X(x_i;\bm{\widehat{\theta}}_m),i=1,\ldots,n\}$
  to obtain $\bm{\widehat{\theta}}_c$. Moreover, in the second step we
  use a variant on the sequential procedure proposed
  by~\cite{bib:aas-czado-frigessi-bakken-09} to select pair copulas
  one by one; we refer to this as \textit{incremental} copula
  inference. The final stage of our method is a joint maximization of~\eqref{eq:24}
  over all model parameters (with the exception of the
  fulcrum $\delta$, for reasons we later explain) using the estimates
  from IFM as starting values. 

  Statistical theory for copula inference for
  time series is a developing area. For first-order Markov
  processes, \cite{bib:chen-fan-06b} showed consistency and asymptotic
  normality of a two-step procedure in which the
  margin is estimated non-parametrically and the copula is estimated
  parametrically. \cite{bib:nagler-krueger-min-20}
  have recently proposed
  multivariate time series models combining parametric margins
  and s-vine copula processes, which subsume our
  models in the case where the v-transform is degenerate; they formulate
  regularity conditions under which the multi-step sequential method
  of~\cite{bib:aas-czado-frigessi-bakken-09} yields
  consistent and asymptotically normal parameter estimates. 

  In the following sections we elaborate first on the critical
  marginal modelling phase before discussing inference for
   the different elements of the copula density $c_{\bm{U}}$.

 \subsection{Marginal modelling}\label{sec:marginal-models}

We are free to choose the best-fitting marginal distributions we can
find for the data. In contrast, well-known econometric models are more constrained in
the marginal behaviour they can model. In particular, many time series
models in the GARCH family have the property that the resulting tails
of the marginal distribution are regularly varying regardless of the
choice of innovation distribution, i.e.~they follow a
power law~\citep{bib:mikosch-starica-00}. However, in real asset return
data we often encounter situations where tail behaviour differs in the
two tails and one or both may be lighter than a regularly-varying law
would dictate.

\begin{figure}[hh]
\centering
\includegraphics[width=14cm,trim=.5cm .5cm .5cm .5cm,clip]{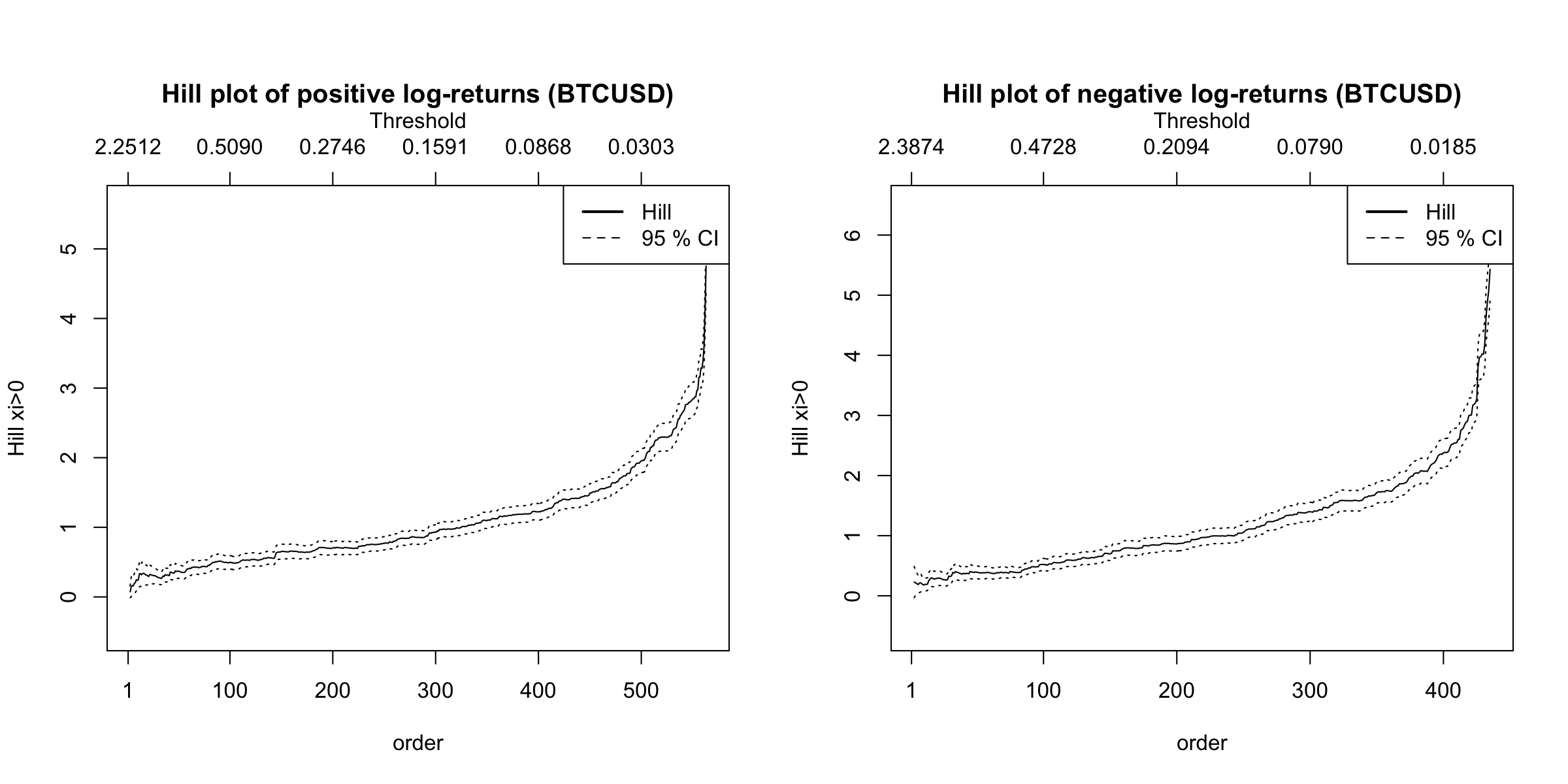}
\includegraphics[width=14cm,trim=.5cm .5cm .5cm .5cm,clip]{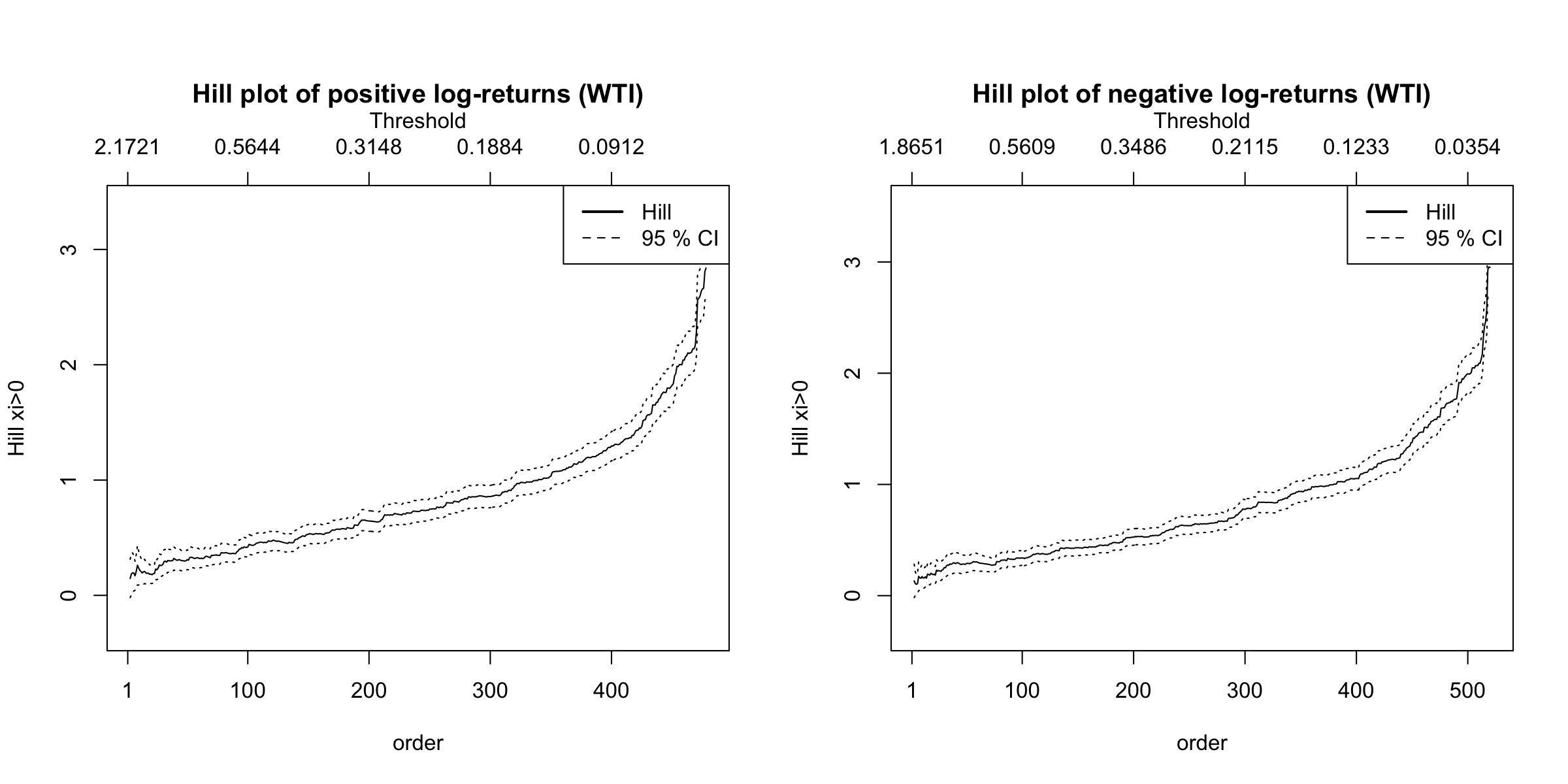}
\includegraphics[width=14cm,trim=.5cm .5cm .5cm .5cm,clip]{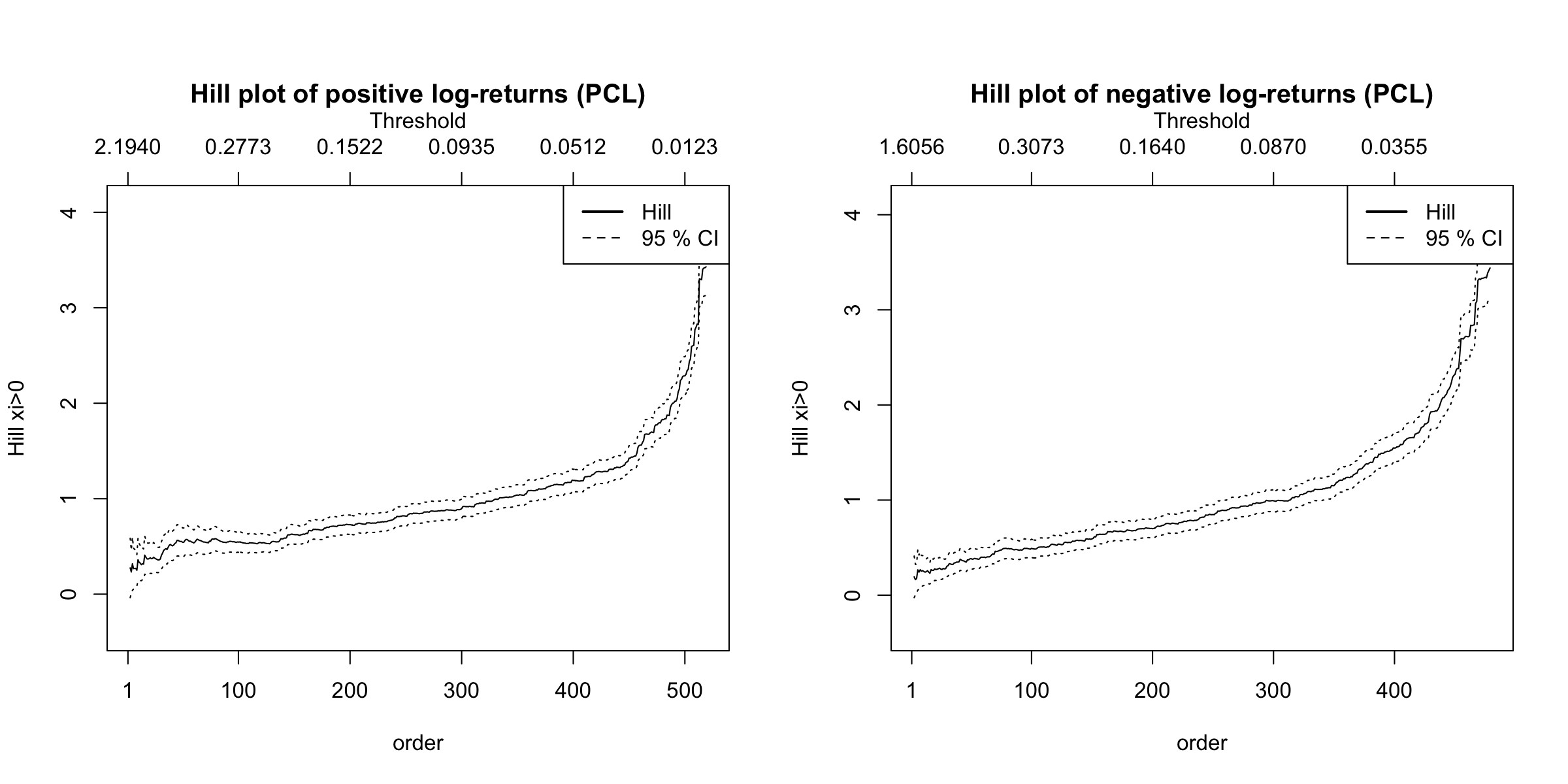}

\caption{Hill plots of the positive and negative log-returns. Top: Bitcoin-USD log-returns (2016-01-01/2019-11-01); middle: WTI crude oil log-returns (2015-01-01/2019-02-12); bottom: PCL log-returns (2006-01-01/2010-01-10).}
\label{Hill_plots}
\end{figure}

See, for instance, Figure \ref{Hill_plots}. For three financial datasets
exhibiting stochastic volatility,
which will be analysed in Section~\ref{sec:empirical-results}, we show
the well-known Hill estimator of the tail index of a regularly-varying
law (see~\cite{bib:hill1975simple} and \cite{bib:de1998asymptotic} for the basic
properties of the Hill estimator). These plots should stabilize
towards the left-hand end at a value
greater than zero if a power tail is justified, but all these plots
appear to continue to decay towards zero.

To model this behaviour, as well as the asymmetry of tails, we
introduce a simple mixture of positive-valued distributions as a model
for marginal distributions. The specific families that we consider are
the generalized gamma distribution and the Burr distribution. In
terms of extreme value theory (EVT) the first of these belongs to the Gumbel domain of attraction, and the second belongs to the Fr\'{e}chet domain of attraction~\citep{bib:embrechts-klueppelberg-mikosch-97}. In both cases, the associated Hill plots have similar shapes to the ones shown in our examples. A convenient feature of the generalized gamma distribution is the fact that for different choices of parameters, it can have tails which are both heavier or lighter than the exponential distribution, and it contains the Weibull distribution as a special case.

In general, we define a two-sided mixture corresponding by taking a
density $f_0(\cdot\,; \bm{\eta})$ on the positive half-line and a
parameter $p\in[0,1]$ and setting
\begin{equation}\label{weibull2_margin}
f\left(x |\,\bm{\eta}^+,\bm{\eta}^{-}\right)=
\left\{\begin{array}{l}
(1-p) \,f_0(-x|\bm{\eta}^{-}); \quad x<0 \\ 
p \,f_0(x|\bm{\eta}^+); \quad x \ge 0.\end{array}\right.
\end{equation}
A mixture of this form can also be
interpreted as the result of splicing
 densities at the origin in a non-continuous
manner\footnote{It is also possible to
splice densities continuously but our results suggest that this
results in slightly inferior fits.}. In the case of the generalized gamma distribution 
$$f_0(x|\, \bm{\eta})=\frac{\nu y^{\nu-1}}{(\mu / \sigma)^{\nu \sigma} \Gamma(\sigma)} y^{\nu(\sigma-1)} \exp \left(-(y \sigma / \mu)^{\nu}\right), \quad \bm{\eta} =(\sigma,\mu,\nu)^\prime>\bm{0},$$ 
while for the Burr distribution 
$$f_0(x|\, \bm{\eta}) = \frac{\alpha \beta (x/\sigma)^\beta}{x
  [1 + (x/\sigma)^\beta]^{\alpha + 1}},\quad
\bm{\eta}=(\alpha,\beta,\sigma)^\prime>\bm{0}.$$ Other choices for $f_0$ such as the
log-gamma distribution can also give good results but the generalized
gamma and Burr are good models for the data we analyse in this paper.

We have also compared these models with various two-sided distributions that are
common in modelling log-returns such as the Student t, skewed Student t
and normal inverse-Gaussian (NIG)
distributions. The mixtures generally give statistically superior
fits.

 \subsection{Incremental copula inference}\label{sec:gener-lagg-data}
 
 In our model, the first term
 $L_2(\bm{\theta}_c;\bm{u})=\log\left(c_{\bm{U}}(u_1\dots,u_n;\bm{\theta}_c)\right)$
 in the log-likelihood~\eqref{eq:24} is maximized with respect to the
 copula parameters $\bm{\theta}_c$
  using pseudo-copula data $\{u_i =
  F_X(x_i;\bm{\widehat{\theta}}_m),i=1,\ldots,n\}$. The term
  $L_2(\bm{\theta}_c;\bm{u})$ further splits into a term coming from
  the copula $c_{W^*}$ in~(\ref{eq:14}) and a term coming from the
  d-vine copula model for $(V_t)$
  which may be written as $ \sum_{i=1}^{k} L_i( \bm{\theta}_{i},
  \bm{\theta}_v;\bm{u})$ where
  \begin{equation}\label{eq:19}
      L_i(
        \bm{\theta}_{i}, \bm{\theta}_v; \bm{u}) =
   \sum_{t=1}^{n-i}  \log \left.  c_i\big(v_{t \mid S_{t,t+i}}
  , \;v_{t+i\mid S_{t,t+i}}; \bm{\theta}_i \big) \right|_{v_1 =
  \vtrans(u_1;\bm{\theta}_v),\ldots,v_n = \vtrans(u_n; \bm{\theta}_v)} 
  \end{equation}
where $\bm{\theta}_{i} = (\theta_1,\ldots,\theta_i)$ and $\theta_i$
denotes the parameter of the $i$th pair copula.
Note that the terms $c_i\big(v_{t \mid S_{t,t+i}}
  , \;v_{t+i\mid S_{t,t+i}}; \bm{\theta}_i \big)$ depend on $\theta_i$
  through the copula $c_i$ and on $\theta_1,\ldots,\theta_{i-1}$ through
 the conditional distributions $v_{t \mid S_{t,t+i}}$ and $v_{t
  +i \mid S_{t,t+i}}$.
We assume to begin
  with that the parameter(s) $\bm{\theta}_v$ of the v-transform are
  fixed and postpone a discussion of their estimation to the next
  section.

 We select pair copulas
from a number of widely-used one-parameter families, these being Gauss, Gumbel,
  Clayton, Frank and Joe, as well as rotations of the Gumbel, Clayton and Joe copulas
  through 180 degrees. We have experimented with non-exchangeable
  generalizations of these copulas using the construction
  of~\cite{bib:liebscher-08}. These break the time reversibility of
  the resulting process but only rarely deliver gains in fit to justify the extra
  complexity and additional parameters. We also omit
  the two-parameter t copula
  from consideration since it adds little in the examples we present and since the restriction to one-parameter copulas permits
  the use of fast sequential method-of-moments estimation in our
  forecasting study.

  In the incremental method we choose the copula
  families for $c_1,\ldots,c_k$ sequentially but we continually
  re-estimate all parameters of previously chosen copulas. We also
  use the incremental method to estimate the order $k$ of the process. The
  algorithm consists of an initialization step and a continuation step:
  \begin{description}
  \item[Step 1:] Select the copula family $c_1$ and parameter value $\theta_1$
    that maximize $L_1 ( \bm{\theta}_{1}, \bm{\theta}_v; \bm{u})$.
    \item[Step $j$:] With copula families $c_1,\ldots,c_{j-1}$ already
  determined, select the copula family $c_j$ and parameter values
  $\bm{\theta}_j$ that maximize $\sum_{i=1}^{j} L_i( \bm{\theta}_{i},
  \bm{\theta}_v;\bm{u})$. If the AIC of the resulting model of order
  $j$ is lower
  than the previous model of order $j-1$, continue; otherwise stop and set $k=j-1$.
  \end{description}
  This differs from the sequential method used
  by~\cite{bib:aas-czado-frigessi-bakken-09} and analysed
  in~\cite{bib:nagler-krueger-min-20} since we do not hold the
  parameters of previously selected copulas fixed at their previously
  estimated values. This should reduce the risk of
  parameter error percolating through the procedure and lead to
  estimates that are even closer to full ML estimates than those
  obtained using the sequential method.
  


  \subsection{Estimating the v-transform}

  A natural approach to estimating the parameters $\bm{\theta}_v$ of the
  v-transform
  is to optimize over these 
  in an outer loop, while applying the incremental procedure of the
  previous section in an inner
  loop. This is feasible, albeit very computationally intensive, since the incremental
  procedure is itself a greedy algorithm.

  Care must be taken in the optimization with respect to the fulcrum
  parameter $\delta$. If $\delta  =u_i =
  F_X(x_i;\bm{\widehat{\theta}}_m)$ for some original data point
  $x_i$, then $\vtrans(u_i) = 0$ and the log-likelihood for the copula
  takes the value $-\infty$. This implies that the profile likelihood  of
  $\delta$ is not differentiable at such points and has multiple local
  maxima. We thus use a grid search for an optimal estimate of
  $\delta$, avoiding the $u_i$ values, rather than continuous optimization in the interval $[0,1]$.
The key observation is that $\delta$ is a \textit{threshold} parameter and
pseudo copula observations $u_i$ on either side lead to different
responses. The situation is akin to that which arises for
TAR and other
threshold models in econometrics; see, for example,~\cite{bib:tong-83}
and~\cite{bib:hansen-00}. In these models it is typical to use
conditional (or profile likelihood) inference for other parameters while the
threshold parameter $\delta$ takes fixed values on a grid.

 In our empirical examples we use a procedure that is simpler on two
 counts, and therefore faster. First, we restrict 
 attention to the linear v-transform. This is sufficient
  to capture the main features of stochastic volatility and outperform
  competitor models in the examples we present; moreover, it
  facilitates value-at-risk (VaR) forecasting and backtesting with the fitted models,
  as we show in Section~\ref{sec:using-model-forec}.
Second, we optimize over the fulcrum
  parameter $\delta$ in the d-vine(1) model of Step 1 above, rather than the full
  d-vine($k$) model.

\subsection{Estimating the copula $C_W$}\label{sec:estim-copula-c_w}

In a gvt-d-vine copula process the log-likelihood contains an extra
additive term which can be fully expressed in terms of 
$C_W(\delta(u_{t-1}),\delta(u_t); \bm{\theta}_W)$ for the bivariate
copula $C_W$ in~\eqref{eq:10}. A final step is added to the stepwise method
in which this additional term is maximized with respect to $\bm{\theta}_W$.

In practice, the extent to which the parameter(s) $\bm{\theta}_W$ of a copula
$C_W$ can be identified from data depends on the extent to which the
set of points $S = \{(\delta(u_{t-1}),\delta(u_t)),\;t=2,\ldots,d\}$ fills
the unit square $[0,1]^2$, which depends in turn upon the choice of
v-transform. 
In the 
special case of the linear v-transform, $S$ is identical to the singleton $\{(\delta,\delta)\}$ meaning that we can only identify the
value of $C_W$ at this point (see the left panel of
Figure~\ref{Wast_copulas}).

For the case of a non-linear v-transform we recommend picking
a copula family $C_W$ that can model positive dependence, negative dependence
and independence; an ideal candidate is the radially
symmetric Frank copula which involves only simple functions and is
quick to evaluate.  For the linear transform we can actually circumvent the
evaluation of $C_W$ entirely. The likelihood can be
reparameterized in terms of $\lambda = C_W(\delta,\delta) = \P(W_{t-1}
\leq \delta, W_t \leq \delta) = \P(U_{t-1}
\leq \delta, U_t \leq \delta)$, or in terms of a correlation
parameter
\begin{equation}
  \label{eq:12}
  \rho_U = \rho(\indicator{U_{t-1} \leq \delta}, \indicator{U_{t} \leq
    \delta}) = \rho(\indicator{W_{t-1} \leq \delta}, \indicator{W_{t} \leq
    \delta}) = \frac{\lambda -
\delta^2}{\delta-\delta^2}.
\end{equation}

\subsection{A graphical method using generalized lagging}\label{sec:graph-meth-using}

We use a graphical method to gain insight into the fitted pair copula
model. 
Having fitted the model, we first construct pseudo-copula data $\{v_i
=
\vtrans(F_X(x_i;\bm{\widehat{\theta}}_m);\bm{\widehat{\theta}}_V),i=1,\ldots,n\}$
from $c_{\bm{V}}$. We then use a
sequential method to reconstruct, for $i= 1,\ldots,k$, the datasets
\begin{equation}\label{eq:23}
\mathcal{D}_i = \left\{\left(v^{(i)}_{1,t},   v^{(i)}_{2,t}
  \right),\;\;t = 1, \ldots, n-i\right\},\quad v^{(i)}_{1,t} = v_{t
  \mid S_{t,t+i}} ,\quad   v^{(i)}_{2,t} = v_{t+i \mid S_{t,t+i}},
\end{equation}
which are modelled by copula $c_i$ in the likelihood contribution~(\ref{eq:19}).
The points of the lag 1 dataset $\mathcal{D}_1$ are given by $\left(v^{(1)}_{1,t} , v^{(1)}_{2,t}\right)  = \left(v_t, v_{t+1}\right)$,
  $t=1,\ldots,n-1$, and these are the points that would be used in a
  first-order lagplot. For $i < k$ the
  recursive formulas~(\ref{eq:18}) then allow the recursive
  construction of dataset $\mathcal{D}_{i+1}$ from $\mathcal{D}_i$ by
  setting
  \begin{displaymath}
    v^{(i+1)}_{1,t} = h_1^{(i)}\left( v^{(i)}_{1,t},
      v^{(i)}_{2,t}\right),\;   v^{(i+1)}_{2,t} = h_2^{(i)}\left(
      v^{(i)}_{1,t+1}, v^{(i)}_{2,t+1}\right),\quad t = 1, \ldots, n-
    i -1.
  \end{displaymath}
The $h$-functions of the previous copula $C_i$ are always required to construct the next dataset
$\mathcal{D}_{i+1}$. We refer to the iterative procedure as
\textit{generalized lagging} since a scatterplot of the points in the dataset
$\mathcal{D}_i$ may be thought of as a kind of lagplot at generalized
lag $i$. In the graphical method we plot sample Kendall's tau values for each dataset
$\mathcal{D}_i$ against the theoretical values for the
fitted copulas $c_i$ to see the degree of correspondence between sample
and fitted measures.

  The generalized lagging procedure also suggests a fast alternative
  to the sequential ML estimation method
  of~\cite{bib:aas-czado-frigessi-bakken-09}. We could use the dataset
  $\mathcal{D}_i$ at lag $i$ to estimate $\theta_i$ by a
  method-of-moments procedure that exploits the one-to-one relationsip
  between $\theta_i$ and Kendall's tau for the copula families of interest.
  This is a much faster method than our main incremental method
  and we use it in a rolling estimation
  and backtesting study at the end of the paper.

    \section{Empirical Results}\label{sec:empirical-results}

    \subsection{Data and models}

In this section we consider three different financial datasets and their
estimation using the models and methods introduced in the previous sections.
The first dataset consists of the log-returns of the
Bitcoin-USD exchange from January 1, 2016 to November 1, 2019. The second dataset comprises the
log-returns of the WTI crude-oil price from January 1, 2015 to February 12, 2019. Finally, the third series consists of the log returns of the price of the PCL stock from January 1, 2006 to January 10, 2010. Each series consists of $1000$
observations\footnote{Note that in the presented results the values of
  log-returns have been multiplied by 10 for
  improved stability in fitting some of the alternative GARCH-type specifications.}.

The first two datasets do not exhibit
strong serial correlation (as is usual for log-returns), but they do
show stochastic volatility. In contrast, both serial correlation and
stochastic volatility are present in the third dataset. The third
dataset was identified by taking all S\&P500 stocks for the
2006--10 time period and selecting those with the largest absolute
values of first-order serial correlation in log-returns. Thus it can be considered as
a more extreme example of the level of serial correlation that is
present in raw log-returns and a good candidate series for exploring
the added value of a gvt-d-vine model over a standard vt-d-vine-model.


In addition to the vt-d-vine model introduced in this paper we also
consider a vt-ARMA model of the type presented
in~\cite{bib:mcneil-20}. To examine the extent to which
  higher-order Markov models are necessary, we also compare with a
  vt-d-vine model of order 2. From the wider GARCH family inspired by the
ideas of~\cite{bib:engle-82} we fit the standard GARCH model
of~\cite{bib:bollerslev-86}, the exponential GARCH model of~\cite{bib:nelson-91} and
the GJR-GARCH model of~\cite{bib:glosten-jagannathan-runkle-93}. We
also vary the choice of innovation distribution for these
models. 
The optimal vt-ARMA and GARCH model orders are selected by minimization of the AIC criterion.


\subsection{Parameter estimates and model comparison}

\begin{table}[ht]
  \centering
 \begin{tabular}{rlll}
 \toprule
 Parameter & BTCUSD & WTI & PCL \\ 
 \midrule
 $\theta_1$ & 1.4310 (frank) & 1.1670 (joe) & 1.3156 (gumbel) \\ 
 $\theta_2$ & 1.2872 (frank) & 0.1260 (gaussian) & 1.1709 (gumbel) \\ 
 $\theta_3$ & 0.8572 (frank) & 1.0338 (gumbel) & 0.2603 (clayton) \\ 
 $\theta_4$ & 0.1614 (clayton) & 1.0368 (joe) & 1.1048 (gumbel) \\ 
 $\theta_5$ & 0.9668 (frank) & 0.0718 (gaussian) & 1.1807 (joe) \\ 
 $\theta_6$ & 0.1046 (clayton) & 1.0504 (joe) & 0.1492 (clayton) \\ 
 $\theta_7$ & 0.0209 (clayton) & 0.0725 (gaussian) & 0.1197 (clayton) \\ 
 $\theta_8$ & 0.0327 (gaussian) &  & 0.1068 (gaussian) \\ 
 $\theta_9$ & 0.3062 (frank) &  & 1.0505 (gumbel) \\ 
 $\theta_{10}$ & 0.5875 (frank) &  & 0.5380 (frank) \\ 
 $\theta_{11}$ &  &  & 0.0737 (clayton) \\ 
 $\theta_{12}$ &  &  & 1.0856 (joe) \\ 
 $\theta_{13}$ &  &  & 0.0678 (clayton) \\ 
 $\delta$ & 0.45 & 0.3 & 0.55 \\ 
 $\eta^+_{1}$ & 3.5505 & 8.9849 & 1.2536 \\ 
 $\eta^+_{2}$ & 0.0447 & 0.0050 & 1.4853 \\ 
 $\eta^+_{3}$ & 0.4516 & 0.3523 & 0.1286 \\ 
 $\eta^{-}_{1}$ & 1.6490 & 1.7268 & 2.6140 \\ 
 $\eta^{-}_{2}$ & 0.1815 & 0.3119 & 1.1650 \\ 
 $\eta^{-}_{3}$ & 0.6380 & 0.9064 & 0.3332 \\
$p_0$ &  0.5510 & 0.4786 & 0.5239\\
 \bottomrule
 \end{tabular}
 \caption{Summary of parameter estimates for vt-d-vine copula models.}
\label{table:estimates}
 \end{table}

 Table~\ref{table:estimates} shows the estimates for the vt-d-vine
 models fitted to the datasets. The parameters of the pair copulas are simply
 denoted $\theta_i$ and the names of the selected copulas are given. The marginal model used for both the vt-processes is a mixture of positive and
negative distributions as in \eqref{weibull2_margin} amounting to $7$ marginal parameters when using the generalized gamma (for the first two datasets) and Burr (for the third dataset) distributions. Marginal parameters are denoted
 $\eta_j^{-}$ and $\eta_j^{+}$ for the negative and positive tails respectively.
The fulcrum estimates are $\delta = 0.45,\, 0.3,\, 0.55$ for the
three datasets. 

Table~\ref{allmodel_summary}
summarizes measures of fit for the best fitting model within each of the
five considered classes with each table relating to a different dataset. We see that vt-d-vine models (and sometimes
also vt-ARMA models) fare favourably against the alternatives from the
GARCH family. It is evident that the vt-d-vine model of
  order 2 is not at all competitive for the BTCUSD and PCL data, but
  it does much better for the WTI dataset, where it gives the second
  lowest AIC model after the vt-d-vine model of order 7.

\begin{table}[!htbp]
\centering
\begin{tabular}{llrrrrr}
  \toprule
 Dataset & Model & LogLik & NumPars & AIC & rank & extra AIC \\ 
  \midrule
 BTCUSD & vt-ARMA(1,1) & -373.54 & 9 & 765.08 & 2 & 8.91 \\ 
  & vt-d-vine(2) & -407.77 & 9 & 833.54 & 6 & 77.37 \\ 
 & vt-d-vine(10) & -361.09 & 17 & 756.18 & 1 & 0.00 \\ 
 & sGARCH(1,1) & -401.08 & 6 & 814.17 & 5 & 57.99 \\ 
 & eGARCH(1,1) & -384.61 & 7 & 783.21 & 3 & 27.04 \\ 
&  gjrGARCH(1,1) & -399.06 & 7 & 812.13 & 4 & 55.95  \\ 
   \midrule
WTI & vt-ARMA(1,1) & -591.34 & 9 & 1200.69 & 3 & 12.41 \\ 
&  vt-d-vine(2) & -588.92 & 9 & 1195.83 & 2 & 7.56 \\
&  vt-d-vine(7) & -580.14 & 14 & 1188.28 & 1 & 0.00 \\ 
&  sGARCH(1,1) & -601.43 & 6 & 1214.86 & 5 & 26.58 \\ 
 & eGARCH(6,6) & -579.28 & 22 & 1202.56 & 4 & 14.28 \\  
 & gjrGARCH(1,1) &  -601.40 & 7 & 1216.79 & 6 & 28.52 \\ 
 \midrule
 PCL & vt-ARMA(1,1) & 159.64 & 10 & -299.29 & 5 & 44.67 \\ 
  & vt-d-vine(2) & 82.51 & 9 & -147.02 & 6 & 196.93 \\ 
 & vt-d-vine(13) & 191.98 & 20 & -343.95 & 1 & 0.00 \\ 
 & sGARCH(1,1) & 172.80 & 6 & -333.61 & 2 & 10.34 \\
 & eGARCH(1,1) & 173.77 & 7 & -333.53 & 3 & 10.42 \\ 
&  gjrGARCH(1,1) & 173.36 & 7 & -332.73 & 4 & 11.23 \\ 
   \bottomrule
\end{tabular}
\caption{Summary of all model fits for all datasets.}
\label{allmodel_summary}
\end{table}


For the PCL data, a gvt-d-vine model incorporating a Frank copula for
the $(W_t)$ process gave a significant improvement over the
best vt-d-vine model according to a likelihood ratio test. In terms of
log-likelihood, it was also
superior to any combined ARMA-GARCH model using the GARCH
specifications of Table  \ref{allmodel_summary} although the AIC value of
the best ARMA-GARCH model does `catch up' with the AIC value of the best
gvt-d-vine model in this case.

 Note that the Markov nature of the d-vine models means that more lags need to be
 considered in this approach to obtain similar behaviour to
 models in the GARCH family which effectively have a `moving-average
 term' for modelling volatility.

\subsection{Graphical analysis of fit}

\begin{figure}[!htbp]
\centering
\includegraphics[width=14cm,trim=0cm 0cm 0cm 0cm,clip]{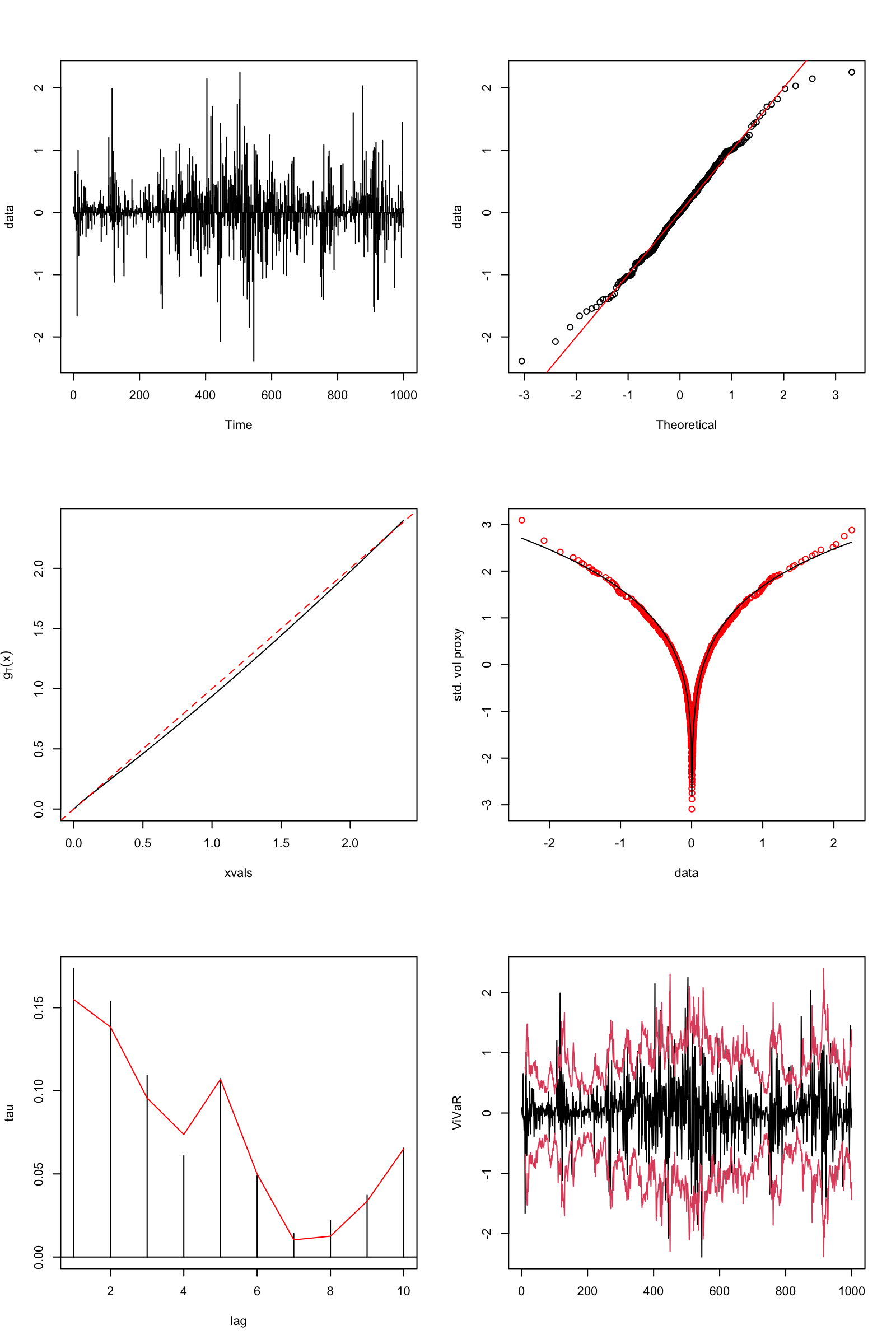}
\caption{Bitcoin data. Top left: log-returns; top right: QQ-plot of
  the fitted versus empirical marginal quantiles; middle: implied
  profile function $g_T$ and standardized volatility proxy; bottom
  left: implied Kendall's tau values from the fitted model (line) and from
  sequential method of moments estimation (bars); bottom right:
  plot of 95\% ViVaR interval.}
\label{btc_gof}
\end{figure}

\begin{figure}[!htbp]
\centering
\includegraphics[width=14cm,trim=0cm 0cm 0cm 0cm,clip]{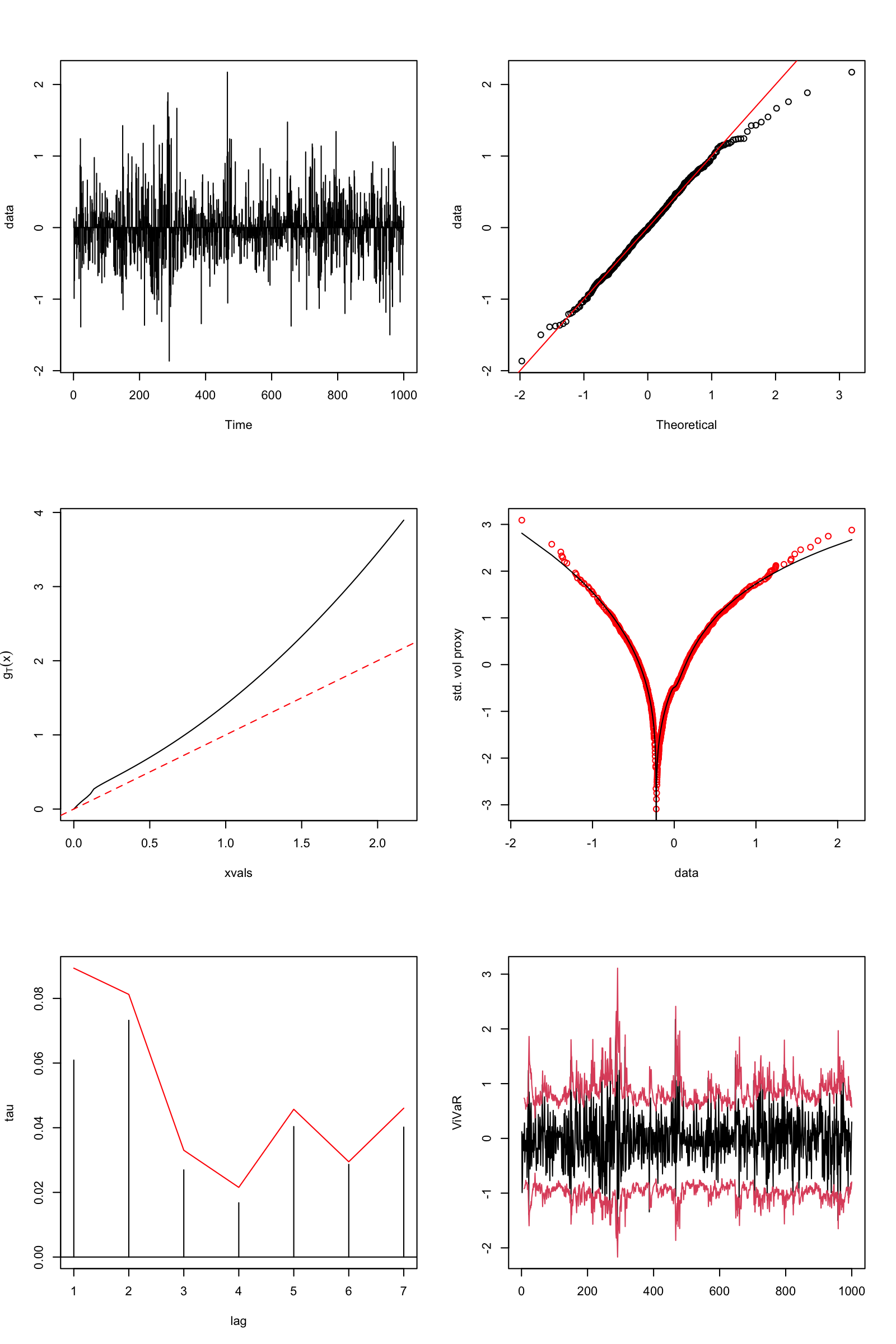}
\caption{WTI data. Top left: log-returns; top right: QQ-plot of the fitted versus empirical marginal quantiles; middle: implied profile function $g_T$ and standardized volatility proxy; bottom
  left: implied Kendall's tau values from the fitted model (line) and from
  sequential method of moments estimation (bars); bottom right:
  plot of 95\% ViVaR interval. }
\label{wti_gof}
\end{figure}

\begin{figure}[!htbp]
\centering
\includegraphics[width=14cm,trim=0cm 0cm 0cm 0cm,clip]{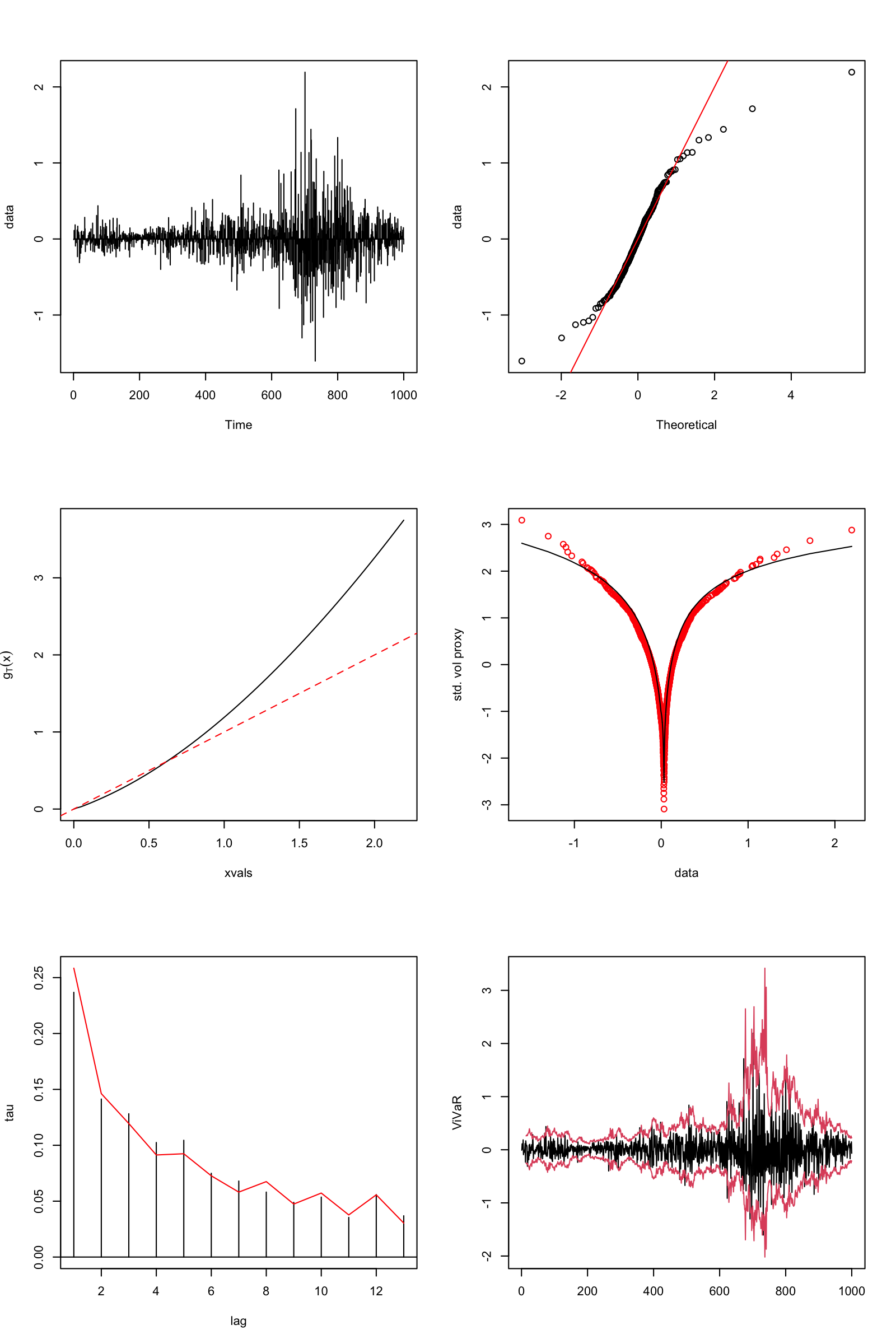}
\caption{PCL data. Top left: log-returns; top right: QQ-plot of the fitted versus empirical marginal quantiles; middle: implied profile function $g_T$ and standardized volatility proxy; bottom
  left: implied Kendall's tau values from the fitted model (line) and from
  sequential method of moments estimation (bars); bottom right:
  plot of 95\% ViVaR interval.}
\label{pcl_gof}
\end{figure}

We assess the fit of the models graphically in Figures \ref{btc_gof}, \ref{wti_gof} and
\ref{pcl_gof}.  In addition to a time series plot of each dataset and
a QQ-plot against the fitted marginal distribution, we show four
other plots. The first of these is a plot of a \textit{partial rank autocorrelation
function} in which the empirical values of
Kendall's tau calculated from the generalized lagged datasets
$\mathcal{D}_i$ in~\eqref{eq:23} are plotted as bars and
the implied Kendall tau values for the corresponding fitted
pair-copulas are plotted as points. 
These plots show good correspondence between data and model values.

The fourth plot for each series shows an estimate of the profile
function $g_T$ of the volatility proxy transformation $T$ as defined
in Section~\ref{sec:v-transforms-uniform}.
The fifth plot shows the volatility proxy
function $T(x) = \Phi^{-1}(\mathcal{V}(F_X(x)))$. The points show an empirical estimate of this
relationship while the curve is the parametric estimate implied by
the fitted model. The final plot is an illustration of an
interval-valued risk measure that we call
$\operatorname{ViVaR}$ which is described in Section~\ref{sec:cond-distr-pred}.

Overall, these sets of plots suggest satisfactory fits. For
the PCL dataset, the curvature of the marginal QQ-plot does suggest a slight
lack of fit in the tails. This seems to be introduced in the final
joint optimization over all parameters and suggests a lack of fit in
the copula component, possibly caused by some remaining misspecification of the
tail dependencies in the copula model. The estimated marginal
distribution
obtained in the first step of the IFM model yields a better
looking QQ-plot (omitted).


The estimated profile function $g_T$ for the Bitcoin data is that of a
symmetric volatility proxy function. Moreover the  smallest value of
the volatility proxy corresponds with a zero return. For the WTI and
PCL data there is more asymmetry. The estimated values of $\mu_T$ are
not equal to zero and both profile functions satisfy $g_T(x) > x$ for
large $x$. This can be interpreted as large negative log-returns
contributing more to the volatility proxy variable.

\section{Using the Model for Forecasting}\label{sec:using-model-forec}

In this section we consider the out-of-sample performance of the
best models from the previous section, looking in particular at predictions of value-at-risk (VaR).

\subsection{Prediction of value-at-risk measures}\label{sec:cond-distr-pred}

In our most general model, conditional quantiles of the forecast distribution $F_{U_t\mid
  U_{t-1}, \ldots, U_{t-k}}$ can be calculated via
numerical integration of the conditional density in
equation~(\ref{cond_dist_gvtcopar}). In the model with linear
v-transform considerable simplification is possible. In this case,
using the shorthand notation $\bm{U}_{t-1} = (U_{t-1},\ldots,U_{t-k})$, the
conditional distribution function satisfies, for
$u_t \leq \delta$,
\begin{align}
  F_{U_t|\bm{U_{t-1}}}(u_t|\bm{u_{t-1}})&= \P\left(U_t \leq u_t \mid
    \bm{U}_{t-1} =\bm{u}_{t-1}  \right)\nonumber \\ &=
     \P \left(V_t \geq \vtrans(u_t)\mid
          \bm{U}_{t-1} = \bm{u}_{t-1} \right)
       \P\left( W_t \leq \delta \mid U_{t-1} =
        u_{t-1} \right), \label{eq:22}
    \end{align}
 where the second term satisfies
    \begin{displaymath}
     \P\left( W_t \leq \delta \mid U_{t-1} =
       u_{t-1} \right) = d(u_{t-1}; \delta, \rho_U) =
     \begin{cases}
     \delta + \rho_U(1-\delta) & \text{if $u_{t-1} \leq \delta$} \\
      \delta(1-\rho_U) & \text{if $u_{t-1} > \delta$}
       \end{cases}
     \end{displaymath}
     where $\rho_U$ is the correlation parameter
     describing serial correlation in the $(U_t)$ process in~(\ref{eq:12}). In the case where $\rho_U=0$ we simply have $d(u_{t-1};\delta,\rho_U) =
     \delta$.
     
To find the $\alpha$-VaR at time $t$ for $\alpha \gg 0.5$
we have to solve the equation
\begin{displaymath}
  \P \left(V_t \geq v_t \mid
          \bm{U}_{t-1} = \bm{u}_{t-1} \right) = \frac{1-\alpha}{d(u_{t-1};\delta,\rho_U)}
\end{displaymath}
for $v_t$ and then compute $\VaR_{t,\alpha} = F_X^{-1}(
\vtrans^{-1}(v_t))$. The calculation is facilitated by the fact that
$V_t$ depends on the past values of
$\bm{U}_{t-1}$ only through the past values of $\bm{V}_{t-1} =
\vtrans(\bm{U}_{t-1})$ and hence
\begin{displaymath}
   \P \left(V_t \geq v_t \mid
          \bm{U}_{t-1} = \bm{u}_{t-1} \right) =\P \left( V_t \geq v_t \mid
                  \bm{V}_{t-1} = \bm{v}_{t-1} \right) =
                  1 - F_{V_t \mid \bm{V}_{t-1}}\left( v_t \mid
        \bm{v}_{t-1} \right) = 1 - v_{t \mid
      S_{t-k-1,t}}
\end{displaymath}
where the final expression uses the notation of
Section~\ref{section-Dvine} and we recall that the conditional probabilities $v_{t \mid
  S_{t-k-1,t}}$ can be calculated recursively.

    While full calculation of VaR is relatively straightforward in
    models with a linear v-transform, the above calculations suggest
    that in general models an alternative measure of risk could be
    considered by concentrating on quantiles of the conditional
    distribution of $V_t$ in the first term of~\eqref{eq:22}. For $\alpha \gg 0.5$
let $v_{\alpha,t}$ be given by $1- \alpha  =  \P \left( V_t \geq v_{\alpha,t} \mid
                  \bm{V}_{t-1} = \bm{v}_{t-1} \right)$.
The event $\{V_t \geq
    v_{\alpha,t}\}$ can be written as
    \begin{eqnarray*}
  \{V_t\geq
    v_{\alpha,t}\}  & = & \left\{ U_t \leq \vtrans^{-1}(v_{\alpha,t})\right\} \cup   \left\{
                          U_t \geq \vtrans^{-1}(v_{\alpha,t}) + v_{\alpha,t}\right\} \\
      &= & \Big\{ X_t \in
           \underbrace{\Big(F_X^{-1}\big(\vtrans^{-1}(v_{\alpha,t})\big) ,
           F_X^{-1}\big(\vtrans^{-1}(v_{\alpha,t}) +
           v_{\alpha,t}\big)  \Big)}_{I_{\alpha,t}} \Big\}^\complement
    \end{eqnarray*}
    and thus the interval labelled $I_{\alpha,t}$
    plays the role of a set-valued risk measure
    with the interpretation that its probability is exactly equal to
    $\alpha$ and all points outside it lead to values of the
    volatility proxy that exceed the
    $\alpha$-quantile of volatility. We refer to $I_{\alpha,t}$
    as a volatility-implied VaR interval (ViVaR) and have calculated it
    for the examples of Section~\ref{sec:empirical-results}.


\subsection{Out-of-sample forecasting experiments}

We evaluate the predictive performance of the vt-d-vine models using a rolling
backtesting procedure typical for financial risk
applications; we construct models using a moving window of $n=500$
data, calculate the one-step-ahead predictive
distribution and then validate the model using
probability-integral-transform (PIT) values and value-at-risk (VaR)
exceptions. Data are taken for a 5-year period; these are the years
2015--19 for BTCUSD and WTI and the years 2006--10 for PCL.

To facilitate backtesting of vt-d-vine models in reasonable time, we use an accelerated 
fitting procedure based on the sequential estimation idea described at
the end of
Section~\ref{sec:graph-meth-using}. We follow the IFM approach and first estimate the marginal
distribution $F_X$. We then fix $k$ so that we do not optimize over model order; $k=5$ is sufficient to
give good results for BTCUSD and WTI, while $k=10$ is required for the PCL dataset. We estimate
parameters using the
sequential method-of-moments procedure based on Kendall's tau. For PCL we restrict attention to vt-d-vine models
without serial dependence in $(W_t)$.

The PIT method follows the general model validation approach of~\cite{bib:diebold-gunther-tay-98}.
Suppose at time $t$ we have data $x_{t-n}, \ldots, x_{t-1}$. We estimate the marginal model
$F_X$ and use this to
obtain transformed data $u_{t-n},\ldots,u_{t-1}$ on the u-scale, which
we use to estimate the copula model.
The conditional cdf is given by~(\ref{eq:22}) and this can be used to
calculate a PIT value $\widehat{u}_t = F_{U_t \mid \bm{U}_{t-1}}(u_t
\mid \bm{u}_{t-1})$ where $u_t = F_X(x_t)$ is calculated from a new
out-of sample observation $x_t$. We refer to the sequence of values
$(\widehat{u}_t)$ as \textit{u-PIT values}; if the forecast models are
reasonable they should be iid uniform. We test uniformity with a
Kolmogorov-Smirnov test and independence using the Wald-Wolfowitz runs test and the
Ljung-Box test based on the first 10 autocorrelation estimates;
$p$-values are denoted $p_{KS}$,
$p_{R}$ and $p_{L10B}$ respectively and given in
Table~\ref{table:pred-PITs}. There is a question mark over the
independence of the u-PIT values for PCL but not their uniformity;
this is explained by the dropping of the model  for $(W_t)$ for
computational speed. The Ljung-Box test also gives a significant
result for the Bitcoin data although this was not confirmed by the
runs test. Otherwise the $p$-values are non-significant.

The u-PIT values reveal very little about the quality of stochastic volatility
modelling which is key to making reasonable predictions of VaR; to
validate the modelling of volatility we consider \textit{v-PIT values} given by $\widehat{v}_t = F_{V_t \mid \bm{V}_{t-1}}(v_t
\mid \bm{v}_{t-1})$ where the values $v_t =
\vtrans_{\delta}(u_t)$ are obtained using the estimated linear
v-transform and where $F_{V_t \mid \bm{V}_{t-1}}$ is the predictive
distribution of the underlying d-vine model as before. The v-PIT values should
also be iid uniform and we run the same tests as for u-PIT values and
show results in Table~\ref{table:pred-PITs}. The tests of uniformity
and independence of v-PIT values are not rejected for any of the time series.

\begin{table}[htbp]
  \centering
  \begin{tabular}{lrrrrrrr} \toprule
    &          &                                      \multicolumn{3}{c}{Tests
                                                   on u-PITs} &   \multicolumn{3}{c}{Tests
                                                   on v-PITs} \\
 Dataset & $n$ & $p_{KS}$ & $p_{R}$ & $p_{LB10}$ & $p_{KS}$
                                          & $p_{R}$ & $p_{LB10}$ \\ \midrule
    BTCUSD & 803 & 0.53 & 0.31 &
                                                                      0.03
                                            & 0.62 & 0.92 & 0.14\\
   WTI & 720 & 0.23 & 0.77 & 0.50 & 0.13 & 0.99 & 0.61 \\
    PCL & 744 & 0.51 & 0.00 & 0.01 & 0.20 & 0.88 & 0.82 \\
    \bottomrule
  \end{tabular}
  \caption{\label{table:pred-PITs} $p$-values for tests for uniformity and independence of PIT values applied
  to u-PIT and v-PIT values; $p_{KS}$,
$p_{R}$ and $p_{LB10}$ refer to Kolmogorov-Smirnov test, runs test and
Ljung-Box test using 10 lags.}
\end{table}

The VaR exception tests are based on comparing a sequence of $\alpha$-VaR estimates
$(\operatorname{VaR}_{t,\alpha})$ at times $t$ with the
corresponding realized values $(x_t)$ and counting exceptions $(x_t <
-\operatorname{VaR}_{t,\alpha})$.
In Table~\ref{table:pred-VaR} we
give the results of a two-sided binomial score test applied to the VaR
exception data for $\alpha = 0.95$ and compare with the other
models. The only significant test results are for the standard GARCH
and GJR-GARCH models applied to the Bitcoin data; in all other cases
the exception counts are in line with the expected number under the
hypothesis that the forecast distributions are correct.

\begin{table}[htbp]
  \centering
  \begin{tabular}{llrrrr} \toprule
 Dataset & Model    & $n$ & $n_e$ & $n_e/n$ & $p_B$ \\ \midrule
    BTCUSD & vt-d-vine(5) & 803 & 48 & 0.060 & 0.20 \\
   & sGARCH(1,1) & 803 & 59 & 0.073 & 0.00 \\
    & eGARCH(1,1) & 803 & 46 & 0.057 & 0.34 \\
         & gjr-GARCH(1,1) & 803 & 62 & 0.077 & 0.00 \\ \midrule 
        WTI & vt-d-vine(5) & 720 & 47 & 0.065 & 0.06 \\
   & sGARCH(1,1) & 720 & 41 & 0.057 & 0.39 \\
    & eGARCH(1,1) & 720 & 47 & 0.065 & 0.06 \\
         & gjr-GARCH(1,1) & 720 & 44 & 0.061 & 0.17 \\ \midrule
    PCL & vt-d-vine(10) & 744 & 44 & 0.059 & 0.25 \\
   & sGARCH(1,1) & 744 & 35 & 0.047 & 0.71 \\
         & eGARCH(1,1) & 744 & 32 & 0.043 & 0.38\\
    & gjr-GARCH(1,1) & 744 & 33 & 0.044 & 0.48 \\
    \bottomrule
  \end{tabular}
  \caption{\label{table:pred-VaR} Results of two-sided binomial score
    test: $n$ is number of trials; $n_e$ number of exceptions and
    $p_{B}$ is the test $p$-value. VaR is estimated at level $\alpha =0.95$.}
\end{table}

\section{Conclusion}\label{sec:conclusion}
The results in this paper and further unreported analyses suggest that many volatile time series of 
log-returns on financial assets can be successfully modelled using d-vine copula processes in
conjunction with v-transforms and asymmetric mixed marginal
distributions. In many cases the in-sample fits and out-of-sample
predictions are 
superior to those
obtained from all the most widely used members of the extended GARCH
family.
It is noteworthy that the best fitting models are higher-order Markov models 
with relatively short memory. In the examples of this paper the orders
are 7, 10 and 13, corresponding to no more than 3 weeks of trading
days.

The class of models in this paper could be extended in a number of
directions. An obvious topic of interest is multivariate time series
models for multiple volatile return series. The m-vine model of \cite{bib:beare-seo-15} and
the recent work on s-vine models by
\cite{bib:nagler-krueger-min-20} suggest directions for
generalizing the vt-d-vine model to the bivariate and multivariate
case.

Remaining in the univariate case, it would be interesting to find
further tractable models for the bivariate process $(V_t,W_t)$ that is used to
construct the process $(U_t)$ under the stochastic inversion operation
of Theorem~\ref{theorem:multivariate-vtransform}. We have restricted
attention to processes $(V_t)$ and $(W_t)$ that are fully independent of
each other, although the requirement of the result is simply that the
variables $V_t$ and $W_t$ are contemporaneously independent for all
$t$. Cross dependencies between the processes at different lags could
be useful for modelling feedback effects
between the signs of log-returns at time $t$ and volatility at future
times $t+k$, or vice versa, although statistical estimation of the
resulting models is likely to be challenging.

Despite the current popularity of vine copula models, there are
two areas where more theoretical work is needed. The first is the study of the
serial dependence or mixing properties of vine-copula-based time
series. The papers of~\cite{bib:beare-10}
and~\cite{bib:longla-peligrad-12}
make important contributions in the
first-order Markov case but mixing and ergodicity results for more
general processes would be valuable, particularly if the conditions
can be checked for concrete combinations of pair copulas.
The second area is statistical inference for
vine-copula-based time series, where the recent paper
of~\cite{bib:nagler-krueger-min-20} is one of the first to
address the challenge of
underpinning widely-used and intuitive stepwise procedures with regularity conditions to
guarantee the usual desirable properties of estimators. 

\section*{Software}

The analyses were carried out using \textsf{R} and the \texttt{tscopula} package (McNeil and
Bladt, 2020) available on CRAN and at
\texttt{https://github.com/ajmcneil/tscopula}. This makes
use of code for vine copulas from the \texttt{rvinecopulib}  package~\citep{bib:nagler-vatter-20}.




\appendix
\numberwithin{equation}{section}
 \section{Proofs}\label{sec:proofs}
\subsection{Proof of
  Proposition~\ref{prop:inverseplustransform}}\label{sec:proof-inverseplustransform}


We first observe that, conditional on $V =v$, we have
    \begin{eqnarray*}
     \left\{\tilde{U} =  U \right\}  & = &\left\{ \svtrans^{-1}(v,W) =
                                            \vtrans^{-1}(v), U =
                                            \vtrans^{-1}(v) \right\} \cup
\left\{ \vtrans^{-1}(v,W) = v + \svtrans^{-1}(v), U =
                                            v+ \vtrans^{-1}(v)
                                           \right\} \\
      & = &\left\{ W \leq \Delta(v), U =
                                            \vtrans^{-1}(v) \right\} \cup
\left\{ W > \Delta(v), U =
                                            v+ \vtrans^{-1}(v)
                                           \right\} 
   \end{eqnarray*}
Hence, by the independence of $W$ and $U$, it follows that
\begin{eqnarray*}
  \P\left( \tilde{U} =  U \mid V = v\right) &= \Downprob(v)^2 + (1-\Downprob(v))^2
\end{eqnarray*}
and integrating over $v$ we obtain
\begin{eqnarray*}
   \P\left( \tilde{U} = U \right) &= 2 \E\left( \Downprob(V)^2\right) + 1
                                     -2\E \left(\Downprob(V) \right)\\
&= 2\left( \var\left( \Downprob(V) \right) + \downprob^2 \right) + 1 -2\downprob \\
&= 2 \var\left( \Downprob(V) \right) + \downprob^2 + (1-\downprob)^2\,.
\end{eqnarray*}

\subsection{Proof of Theorem~\ref{theorem:multivariate-vtransform}}\label{sec:proof-multivariate-vtransform}
  We will need the fact that
  \begin{equation}
    \label{eq:8}
    p_{\delta,u}(\delta(u)) =
 \frac{(-1)^{\indicator{u \leq \delta}}}{\vtrans^\prime(u)} \quad \text{for all $u \in [0,1]$}
  \end{equation}
 which we first prove. When $u \leq \delta$ we clearly have
  that $p_{\delta,u}(\delta(u)) = \delta(u) = \Delta(\vtrans(u)) =
  -1/\vtrans^\prime(u)$ where we have used equation~(\ref{eq:7}).
  When $u > \delta$ we have
  that $p_{\delta,u}(\delta(u)) = 1 - \Delta(\vtrans(u)) =
  1  + 1/\vtrans^\prime(u^*)$ where $u^*$ is the dual point of
  $u_i$ as defined in Definition~\ref{def:v-transform}. We have to
  show that $ 1  + 1/\vtrans^\prime(u^*) = 1/\vtrans^\prime(u)$. This follows from the fact that for any $u >
  \delta$
  with dual point $u^* = u - \vtrans(u)$ we have
  $\vtrans(u) = \vtrans(u^*) = \vtrans(u - \vtrans(u))$ so that
  $\vtrans^\prime(u) = \vtrans^\prime(u^*) (1 - \vtrans^\prime(u))$
  and hence $\vtrans^\prime(u) = \vtrans^\prime(u^*) / (1 +
  \vtrans^\prime(u^*))$.
  
Now fix the point $(u_1,\ldots,u_d) \in [0,1]^d$ and consider the set of events $A_i(u_i)$ defined by
\begin{equation*}
A_i(u_i) =
\begin{cases}
\left\{ U_i \leq u_i\right\}& \text{if $u_i \leq \downprob$} \\ 
\left\{ U_i > u_i\right\}& \text{if $u_i > \downprob$} 
\end{cases}
\end{equation*}
The probability $\P(A_1(u_1),\ldots,A_d(u_d))$ is the probability of
the orthant defined by the point $(u_1,\ldots,u_d)$ and the copula 
density at this point is given by
\begin{equation}\label{eq:4}
c_{\bm{U}}(u_1,\ldots,u_d) = (-1)^{\sum_{i=1}^d \indicator{u_i > \downprob}}\frac{\rd^d}{\rd u_1 \cdots \rd u_d} \P\left(\bigcap_{i=1}^d A_i(u_i)\right)\;\;.
\end{equation}
The event $A_i(u_i)$ can be written
\begin{equation*}
A_i(u_i) =
\begin{cases}
\left\{ V_i \geq \vtrans(u_i),\; W_i \leq \Delta(V_i) \right\}& \text{if $u_i \leq \downprob$} \\ 
\left\{ V_i > \vtrans(u_i),\; W_i > \Delta(V_i) \right\}& \text{if $u_i > \downprob$} 
\end{cases}
\end{equation*}
and hence the probability of the
event $\bigcap_{i=1}^d A_i(u_i)$ can be written
\begin{displaymath}
  \int_{\vtrans(u_1)}^1 \cdots \int_{\vtrans(u_d)}^1
  \int_{I_{\delta,u_1}(\Delta(v_1))}\cdots
  \int_{I_{\delta,u_d}(\Delta(v_d))}c_{\bm{V},\bm{W}}(v_1,\dots,v_d,\,
  z_1,\dots,z_d) \rd v_1 \cdots \rd v_d
  \rd z_1\cdots \rd z_d\,.
\end{displaymath}
Taking the derivative of this expression with respect to
$u_1,\ldots,u_d$ and using~(\ref{eq:8}) and~(\ref{eq:4})
yields~\eqref{eq:9}.

When $\bm{V}$ and $\bm{W}$ are independent the joint density factorizes
and~\eqref{eq:15} follows by noting that
\begin{displaymath}
  \left\{ W_i \in I_{\delta,u}(\delta(u)) \right\} =
  \begin{cases}
    \{W_i \leq \delta(u)\} & \text{if $u \leq \delta$}\\
    \{ 1- W_i \leq 1 - \delta(u)\} & \text{if $u > \delta$}
  \end{cases}
   \quad
   = \left\{p_{\delta,u}(W_i) \leq p_{\delta,u}(\delta(u))\right\}\;.
\end{displaymath}
Clearly~(\ref{eq:15}) reduces to the simple form
 $c_{\bm{V}}(\vtrans(u_1),\ldots,\vtrans(u_d))$ when $C_{\bm{W}}$ is the
    independence copula. 
 

    \subsection{Proof of Proposition~\ref{prop:Markov-conditioning}}\label{sec:proof:Markov-conditioning}
    The decomposition of $c_{\bm{W}^\ast}$ into a product of bivariate
    densities $c_{W^\ast}$ is a straightforward consequence of the
    first-order Markov assumption. We then have that
    \begin{displaymath}
      c_{W^\ast}(u_1, u_2) = \frac{C_{p_{\delta,\bm{u}}(\bm{W})}\Big(p_{\delta,u_1}\big(\delta(u_1)\big),
    p_{\delta,u_2}\big(\delta(u_2)\big)\Big)
  }{p_{\delta,u_1}\big(\delta(u_1)\big)
    p_{\delta,u_2}\big(\delta(u_2)\big)} = \frac{\P\left(
                                                        p_{\delta,u_1}(W_1)
                                                        \leq
                                                        p_{\delta,u_1}\big(\delta(u_1)\big),
                                                        p_{\delta,u_2}(W_2)
                                                        \leq
                                                        p_{\delta,u_2}\big(\delta(u_2)\big)   \right) }{p_{\delta,u_1}\big(\delta(u_1)\big)
    p_{\delta,u_2}\big(\delta(u_2)\big)}
    \end{displaymath}
and~(\ref{eq:10}) follows easily by considering the four possible
combinations of values for $p_{\delta,u_1}(\cdot)$ and $p_{\delta,u_2}(\cdot)$.
    

\bibliographystyle{apalike}
\setcitestyle{authoryear,open={(},close={)}}

\begin{thebibliography}{}

\bibitem[Aas et~al., 2009]{bib:aas-czado-frigessi-bakken-09}
Aas, K., Czado, C., Frigessi, A., and Bakken, H. (2009).
\newblock Pair-copula constructions of multiple dependence.
\newblock {\em Insurance: Mathematics and Economics}, 44(2):182--198.

\bibitem[Beare, 2010]{bib:beare-10}
Beare, B. (2010).
\newblock Copulas and temporal dependence.
\newblock {\em Econometrica}, 78(395--410).

\bibitem[Beare and Seo, 2015]{bib:beare-seo-15}
Beare, B. and Seo, J. (2015).
\newblock Vine copula specifications for stationary multivariate {Markov}
  chains.
\newblock {\em Journal of Time Series Analysis}, 36:228--246.

\bibitem[Bedford and Cooke, 2001a]{bib:bedford-cooke-01}
Bedford, T. and Cooke, R. (2001a).
\newblock {\em Probabilistic Risk Analysis: Foundations and Methods}.
\newblock Cambridge University Press, Cambridge.

\bibitem[Bedford and Cooke, 2001b]{bib:bedford-cooke-01b}
Bedford, T. and Cooke, R.~M. (2001b).
\newblock probability density decomposition for conditionally independent
  random variables modeled by vines.
\newblock {\em Annals of Mathematics and Artificial Intelligence}, 32:245--268.

\bibitem[Bedford and Cooke, 2002]{bib:bedford-cooke-02}
Bedford, T. and Cooke, R.~M. (2002).
\newblock Vines--a new graphical model for dependent random variables.
\newblock {\em Annals of Statistics}, 30(4):1031--1068.

\bibitem[Bollerslev, 1986]{bib:bollerslev-86}
Bollerslev, T. (1986).
\newblock Generalized autoregressive conditional heteroskedasticity.
\newblock {\em Journal of Econometrics}, 31:307--327.

\bibitem[Bradley, 2005]{bib:bradley-05}
Bradley, R. (2005).
\newblock Basic properties of strong mixing conditions: a survey and some open
  questions.
\newblock {\em Probability Surveys}, 2:107--144.

\bibitem[Brechmann and Czado, 2015]{bib:brechmann-christian-czado-15}
Brechmann, E.~C. and Czado, C. (2015).
\newblock Copar—multivariate time series modeling using the copula
  autoregressive model.
\newblock {\em Applied Stochastic Models in Business and Industry},
  31(4):495--514.

\bibitem[Chen and Fan, 2006]{bib:chen-fan-06b}
Chen, X. and Fan, Y. (2006).
\newblock Estimation of copula-based semiparametric time series models.
\newblock {\em Journal of Econometrics}, 130(2):307--335.

\bibitem[Chen et~al., 2009]{bib:chen-wu-yi-09}
Chen, X., Wu, W.~B., and Yi, Y. (2009).
\newblock Efficient estimation of copula-based semiparametric {Markov} models.
\newblock {\em Annals of Statistics}, 37(6B):4214--4253.

\bibitem[Creal et~al., 2013]{bib:creal-koopman-lucas-13}
Creal, D., Koopman, S., and Lucas, A. (2013).
\newblock Generalized autoregressive score models with applications.
\newblock {\em Journal of Applied Econometrics}, 28:777--795.

\bibitem[Darsow et~al., 1992]{bib:darsow-nguyen-olsen-92}
Darsow, W., Nguyen, B., and Olsen, E. (1992).
\newblock Copulas and {Markov} processes.
\newblock {\em Illinois Journal of Mathematics}, 36(4):600--642.

\bibitem[De~Haan and Resnick, 1998]{bib:de1998asymptotic}
De~Haan, L. and Resnick, S. (1998).
\newblock On asymptotic normality of the hill estimator.
\newblock {\em Stochastic Models}, 14(4):849--866.

\bibitem[Diebold et~al., 1998]{bib:diebold-gunther-tay-98}
Diebold, F., Gunther, T., and Tay, A. (1998).
\newblock Evaluating density forecasts with applications to financial risk
  management.
\newblock {\em International Economic Review}, 39(4):863--883.

\bibitem[Ding et~al., 1993]{bib:ding-engle-granger-93}
Ding, Z., Granger, C.~W., and Engle, R.~F. (1993).
\newblock A long memory property of stock market returns and a new model.
\newblock {\em Journal of Empirical Finance}, 1:83--106.

\bibitem[Domma et~al., 2009]{bib:domma-giordano-perri-09}
Domma, F., Giordano, S., and Perri, P.~F. (2009).
\newblock Statistical modeling of temporal dependence in financial data via a
  copula function.
\newblock {\em Communications if Statistics: Simulation and Computation},
  38(4):703--728.

\bibitem[Embrechts et~al., 1997]{bib:embrechts-klueppelberg-mikosch-97}
Embrechts, P., Kl\"uppelberg, C., and Mikosch, T. (1997).
\newblock {\em Modelling Extremal Events for Insurance and Finance}.
\newblock Springer, Berlin.

\bibitem[Engle, 1982]{bib:engle-82}
Engle, R.~F. (1982).
\newblock Autoregressive conditional heteroskedasticity with estimates of the
  variance of {United Kingdom} inflation.
\newblock {\em Econometrica. Journal of the Econometric Society}, 50:987--1008.

\bibitem[Fan and Patton, 2014]{bib:fan-patton-14}
Fan, Y. and Patton, A. (2014).
\newblock Copulas in econometrics.
\newblock {\em Annual Review of Economics}, 6:179--200.

\bibitem[Glosten et~al., 1993]{bib:glosten-jagannathan-runkle-93}
Glosten, L.~R., Jagannathan, R., and Runkle, D.~E. (1993).
\newblock On the relation between the expected value and the volatility of the
  nominal excess return on stocks.
\newblock {\em The Journal of Finance}, 48(5):1779--1801.

\bibitem[Haff et~al., 2010]{bib:haff-aas-frigessi-10}
Haff, H., Aas, K., and Frigessi, A. (2010).
\newblock On the simplified pair copula construction - simply useful or too
  simplistic?
\newblock {\em Journal of Multivariate Analysis}, 101:1296--1310.

\bibitem[Hansen, 2000]{bib:hansen-00}
Hansen, B. (2000).
\newblock Sample splitting and threshold estimation.
\newblock {\em Econonetrica}, 68(3):575--603.

\bibitem[Hill, 1975]{bib:hill1975simple}
Hill, B.~M. (1975).
\newblock A simple general approach to inference about the tail of a
  distribution.
\newblock {\em The annals of statistics}, pages 1163--1174.

\bibitem[Ibragimov, 2009]{bib:ibragimov-09}
Ibragimov, R. (2009).
\newblock Copula-based characterizations for higher-order {Markov} processes.
\newblock {\em Econometric Theory}, 25(819--846).

\bibitem[Joe, 1996]{bib:joe-96}
Joe, H. (1996).
\newblock Families of m-variate distributions with given margins and m (m-1)/2
  bivariate dependence parameters.
\newblock {\em Lecture Notes-Monograph Series}, pages 120--141.

\bibitem[Joe, 1997]{bib:joe-97}
Joe, H. (1997).
\newblock {\em Multivariate Models and Dependence Concepts}.
\newblock Chapman \& Hall, London.

\bibitem[Kurowicka and Cooke, 2006]{bib:kurowicka-cooke-06}
Kurowicka, D. and Cooke, R. (2006).
\newblock {\em Uncertainty Analysis with High Dimensional Dependence
  Modelling}.
\newblock Wiley, Chichester.

\bibitem[Liebscher, 2008]{bib:liebscher-08}
Liebscher, E. (2008).
\newblock Construction of asymmetric multivariate copulas.
\newblock {\em Journal of Multivariate Analysis}, 99:2234--2250.

\bibitem[Loaiza-Maya et~al., 2018]{bib:louaiza-maya-et-al-18}
Loaiza-Maya, R., Smith, M., and Maneesoonthorn, W. (2018).
\newblock Time series copulas for heteroskedastic data.
\newblock {\em Journal of Applied Econometrics}, 33:332--354.

\bibitem[Longla and Peligrad, 2012]{bib:longla-peligrad-12}
Longla, M. and Peligrad, M. (2012).
\newblock Some aspects of modeling dependence in copula-based {Markov} chains.
\newblock {\em Journalof Multivariate Analysis}, 111:234--240.

\bibitem[McNeil, 2021]{bib:mcneil-20}
McNeil, A. (2021).
\newblock Modelling volatility with v-transforms and copulas.
\newblock {\em Risks}, 9(1):14.

\bibitem[Mikosch and St\u{a}ric\u{a}, 2000]{bib:mikosch-starica-00}
Mikosch, T. and St\u{a}ric\u{a}, C. (2000).
\newblock Limit theory for the sample autocorrelations and extremes of a
  {GARCH(1,1)} process.
\newblock {\em The Annals of Statistics}, 28:1427--1451.

\bibitem[Mroz et~al., 2021]{bib:mroz-fuchs-trutschnig-21}
Mroz, T., Fuchs, S., and Trutschnig, W. (2021).
\newblock How simplifying and flexible is the simplifying assumption in pair
  copula constructions - analytic answers in dimension three and a glimpse
  beyond.
\newblock {\em Electronic Journal of Statistics}, 15(1):1951--1992.

\bibitem[Nagler et~al., 2020]{bib:nagler-krueger-min-20}
Nagler, T., Kr\"uger, D., and Min, A. (2020).
\newblock Stationary vine copula models for multivariate time series.
\newblock Working paper.

\bibitem[Nagler and Vatter, 2020]{bib:nagler-vatter-20}
Nagler, T. and Vatter, T. (2020).
\newblock rvinecopulib: high performance algorithms for vine copula modeling.
\newblock R package version 0.5.5.1.1.

\bibitem[Nelson, 1991]{bib:nelson-91}
Nelson, D.~B. (1991).
\newblock Conditional heteroskedasticity in asset returns: {A}~new approach.
\newblock {\em Econometrica}, 59:347--370.

\bibitem[Patton, 2012]{bib:patton-12}
Patton, A. (2012).
\newblock A review of copula models for economic time series.
\newblock {\em Journal of Multivariate Analysis}, 110:4--18.

\bibitem[Smith, 2015]{bib:smith-15}
Smith, M. (2015).
\newblock Copula modelling of dependence in multivariate time series.
\newblock {\em International Journal of Forecasting}, 31:815--833.

\bibitem[Smith et~al., 2010]{bib:smith-min-almeida-czado-10}
Smith, M., Min, A., Almeida, C., and Czado, C. (2010).
\newblock {Modeling Longitudinal Data Using a Pair-Copula Decomposition of
  Serial Dependence}.
\newblock {\em Journal of the American Statistical Association},
  105(492):1467--1479.

\bibitem[Spanhel and Kurz, 2019]{bib:spanhel-kurz-19}
Spanhel, F. and Kurz, M. (2019).
\newblock Simplified vine copula models: approximations based on the
  simplifying assumption.
\newblock {\em Electronic Journal of Statistics}, 13(1):1254--1291.

\bibitem[St\"ober et~al., 2013]{bib:stoeber-joe-czado-13}
St\"ober, J., Joe, H., and Czado, C. (2013).
\newblock Simplified pair copula extensions---limitations and extensions.
\newblock {\em Journal of Multivariate Analysis}, 119:101--118.

\bibitem[Tong, 1983]{bib:tong-83}
Tong, H. (1983).
\newblock {\em Threshold models in nonlinear time series analysis}.
\newblock Number~21 in Lecture Notes in Statistics. Springer, Berlin.

\bibitem[Zhao et~al., 2018]{bib:zhao-shi-zhang-18}
Zhao, Z., Shi, P., and Zhang, Z. (2018).
\newblock Modeling multivariate time series with copula-linked univariate
  d-vines.
\newblock Preprint on arXiv.

\end{thebibliography}
\newcommand{\noopsort}[1]{}

\end{document}